\documentclass[sigconf]{acmart}
\AtBeginDocument{%
  }

\copyrightyear{2026}
\acmYear{2026}
\setcopyright{cc}
\setcctype{by}
\acmConference[CHI '26]{Proceedings of the 2026 CHI Conference on Human Factors in Computing Systems}{April 13--17, 2026}{Barcelona, Spain}
\acmBooktitle{Proceedings of the 2026 CHI Conference on Human Factors in Computing Systems (CHI '26), April 13--17, 2026, Barcelona, Spain}
\acmPrice{}
\acmDOI{10.1145/3772318.3791210}
\acmISBN{979-8-4007-2278-3/2026/04}

\usepackage{xspace}
\usepackage{enumitem}
\usepackage{array}
\usepackage{multirow}
\usepackage{subcaption}
\usepackage{enumitem}
\usepackage{subcaption}
\usepackage[nohyperlinks]{acronym}

\acrodef{XR}{Extended Reality}
\acrodef{AR}{Augmented Reality}
\acrodef{VR}{Virtual Reality}
\acrodef{IA}{Immersive Analytics}
\acrodef{CSI}{Crime Scene Investigation}
\acrodef{CSR}{Crime Scene Reconstruction}
\acrodef{SR}{Scene Re-enactment}
\acrodef{UI}{User Interface}
\acrodef{HCI}{Human-Computer Interaction}
\newcommand{\XR}{\ac{XR}\xspace}
\newcommand{\AR}{\ac{AR}\xspace}
\newcommand{\VR}{\ac{VR}\xspace}

\newcommand{\CSI}{\ac{CSI}\xspace}
\newcommand{\CSR}{\ac{CSR}\xspace}
\newcommand{\SR}{\ac{SR}\xspace}
\newcommand{\UI}{\ac{UI}\xspace}
\newcommand{\HCI}{\ac{HCI}\xspace}

\newcommand{\etal}{\textit{et~al.}}
\newcommand{\eg}{e.g.,~}
\newcommand{\ie}{i.e.,~}

\usepackage[dvipsnames]{xcolor}

\definecolor{heatRed}{HTML}{D73027}
\definecolor{heatYellow}{HTML}{FFFFBF}
\definecolor{heatGreen}{HTML}{1A9850}
\definecolor{gridBorder}{HTML}{333333}

\usepackage{tikz}

\newcommand{\drawHeatmap}[2]{%
    \begin{tikzpicture}[baseline=(current bounding box.center)]
        \def\cellW{0.5}
        \def\cellH{#1}
        
        \foreach \val [count=\i from 0] in {#2} {
            
            \pgfmathsetmacro{\col}{mod(\i, 3)}
            \pgfmathsetmacro{\row}{int(\i / 3)}
            
            \def\fillCol{white}
            \ifnum\val=-1 \def\fillCol{heatRed} \fi
            \ifnum\val=0  \def\fillCol{gray} \fi
            \ifnum\val=1  \def\fillCol{BlueGreen} \fi

            \filldraw[fill=\fillCol, draw=gridBorder, line width=0.1mm] 
                (\col * \cellW, -\row * \cellH) rectangle ++(\cellW, -\cellH);
        }
    \end{tikzpicture}%
}

\begin{document}

\title[Criminator: An Easy-to-Use XR ``Crime Animator'']{Criminator: An Easy-to-Use XR ``Crime Animator'' for Rapid Reconstruction and Analysis of Dynamic Crime Scenes}

\author{Vahid Pooryousef}
\orcid{0000-0003-4258-4502}
\affiliation{%
  \institution{Department of Human-Centred Computing, Monash University}
  \city{Melbourne}
  \country{Australia}
}
\email{vahid.pooryousef@monash.edu}

\author{Lonni Besançon}
\orcid{0000-0002-7207-1276}
\affiliation{%
  \institution{Department of Science and Technology, Linköping University}
  \city{Norrköping}
  \country{Sweden}}
\email{lonni.besancon@gmail.com}

\author{Maxime Cordeil}
\orcid{0000-0002-9732-4874}
\affiliation{%
  \institution{School of Electrical Engineering and Computer Science, The University of Queensland}
  \city{Brisbane}
  \country{Australia}}
\email{m.cordeil@uq.edu.au}

\author{Chris Flight}
\orcid{0009-0007-7997-3321}
\affiliation{%
  \institution{Forensic Services Department, Victoria Police}
  \city{Melbourne}
  \country{Australia}}
\email{christopher.flight@police.vic.gov.au}

\author{Alastair M Ross AM}
\orcid{0000-0002-6919-1250}
\affiliation{%
  \institution{Department of Forensic Medicine, Monash University}
  \city{Melbourne}
  \country{Australia}}
\email{aliross.ross@monash.edu}

\author{Richard Bassed}
\orcid{0000-0001-5473-055X}
\affiliation{%
  \institution{Department of Forensic Medicine, Monash University}
  \institution{Victorian Institute of Forensic Medicine}
  \city{Melbourne}
  \country{Australia}}
\email{richard.bassed@monash.edu}

\author{Tim Dwyer}
\orcid{0000-0002-9076-9571}
\affiliation{%
  \institution{Department of Human-Centred Computing, Monash University}
  \city{Melbourne}
  \country{Australia}}
\email{tim.dwyer@monash.edu}

\renewcommand{\shortauthors}{Pooryousef et al.}

\begin{abstract}
  Law enforcement authorities are increasingly interested in 3D modelling for virtual crime scene reconstruction, enabling offline analysis without the cost and contamination risk of on-site investigation. Past work has demonstrated spatial relationships through static modelling but validating the sequence of events in dynamic scenarios is crucial for solving a case.
  Yet, animation tools are not well suited to crime scene reconstruction, and complex for non-experts in 3D modelling/animation.
  Through a co-design process with criminology experts, we designed ``Criminator''---a methodological framework and XR tool that simplifies animation authoring. We evaluated this tool with participants trained in criminology (n=6) and untrained individuals (n=12). 
  Both groups were able to successfully complete the character animation tasks and provided high usability ratings for observation tasks. Criminator has potential for hypothesis testing, demonstration, sense-making, and training. Challenges remain in how such a tool fits into the entire judicial process, with questions about including animations as evidence.
\end{abstract}



\begin{teaserfigure}
  \includegraphics[width=\textwidth]{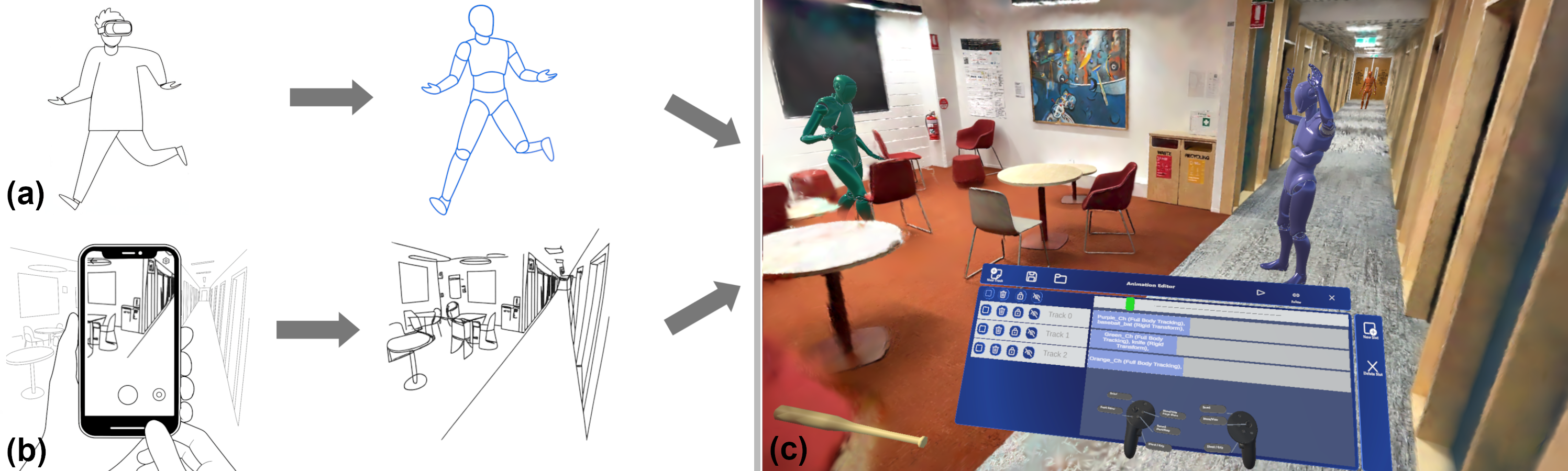}
  \caption{Authoring animated crime scene in VR. (a) creating character animation using a VR headset. (b) 3D scanning a scene using a phone. (c) combination of multiple characters in a 3D scanned environment with an animation editor panel.}
  \Description{This teaser figure illustrates the process of authoring animations in virtual reality, shown in three stages a, b and c: a) A person wearing a virtual reality headset performs actions, which are captured and represented as an animated avatar. b) A crime scene is scanned using a mobile phone to create a digital 3D reconstruction. c) The avatar performance (from a) is combined with the 3D scene (from b) to generate an animated crime scene for analysis and presentation. There are also a floating user interface with virtual reality controllers in c that a user can play and record animation.}
  \label{fig:teaser}
\end{teaserfigure}


\maketitle

\section{Introduction}

In modern criminology, solving a crime requires a multifaceted process that may involve experts from various departments within law enforcement as well as coordination between law and justice agencies. Recent \HCI research has found opportunities for \XR for autopsy, exploring novel techniques for immersive dissection of CT-scan imagery~\cite{affolter_applying_2019,pooryousef2023working,pooryousef2024AutopsyDoc}.
While autopsy focuses on cause of death investigation of the victim, crime examination begins with \CSI, followed by forensic analysis of evidence, and can be further followed by the \CSR process to determine various aspects of criminal events. In this paper, we explore novel opportunities for easy-to-use crime reconstruction in \XR.  
% While current technologies support the determination through analysis and representation, they often rely on complex interfaces and static visualisations, which make them less usable and limit their ability to represent the dynamic aspects of a criminal event that require further exploration.

Based on the definitions provided by OSAC Standard Criteria for \CSR \cite{OSAC2025-N-0004}, 
\CSI refers to an examination of a scene to locate, document, process, collect, and preserve items of potential evidentiary value. On the other hand, \CSR is the process of gaining explicit knowledge of the series of events that surround a scene using deductive and inductive reasoning, physical evidence, scientific processes, and their interrelationships. \CSR deals with particular moments in time and is supported by specific data but often there are gaps between these defined static events.  A \SR is the demonstration of presumed events that can be based, in part, on either physical or testimonial evidence.  \SR is sometimes used to present what may have occurred based on the conclusions of \CSR and can be used to fill in gaps between specific defined events supported by specific data which has been documented or cited.  This often requires a presumption of activity, or actions based on expectations and/or personal influence.  

We refer to \CSR in \VR throughout this paper, but our focus is closer to this latter definition of \SR .
The ultimate goal of \CSI and \CSR processes is the preparation of reports for legal authorities to aid in crime determination. However, conducting a thorough and accurate investigation often requires multiple and/or prolonged visits to the scene, which can risk contamination and evidence loss or destruction. Additionally, crime scenes may be difficult to access or pose safety hazards to investigators~\cite{streefkerk_art_2013}.

In such cases, \CSR enables the further analysis of collected evidence away from the original site, thereby enhancing the comprehensiveness and accuracy of findings without contaminating the scene. This process has been significantly advanced through the use of computer graphics tools and 3D scanners which have become standard practice in \CSR.

Analysis of 3D models of crime scenes is often done using standard desktop workstations with 2D displays.
Such use of technology is typically reserved for high profile cases, due to the high cost of special equipment and skilled modellers. For example, NSW Police's 3D reconstruction of the 2014 Sydney Lindt Caf\'{e} siege was presented in court as a video simulation~\cite{9news2016}.  This simulation revealed the snipers' field of view from multiple angles, enhancing the understanding and assessment of police responses.

Despite the potential for 3D modelling to assist with crime-scene reconstruction, the limited screen size and field of view as well as the lack of depth perception and difficulty of navigating a 3D scene using 2D display and input devices reduces their usability in this role. By contrast, immersive visualisation through \VR and \AR (collectively \XR) technologies enables rendering and experiencing large-scale 3D environments at their actual size, which improves sense of presence and spatial perception. Furthermore, immersive technologies provide more intuitive and efficient interfaces that enables seamless interaction with 3D models, as well as facilitated communication between users. These advantages have motivated many research studies on reconstruction, analysis, and presentation of crime scenes~\cite{streefkerk_art_2013,tolstolutsky_experience_2021,reichherzer_bringing_2021}. 

In addition, existing tools and studies in this domain mostly consider only static representations of crime scenes. While these help to investigate spatial relationships between objects and events, they do not support investigation of temporal relationships and their reconstruction from witness statements or surveillance camera footage. 
Dynamic scenes have been reconstructed for special cases but advanced 3D modelling and animation techniques are required. The potential benefits of \XR technologies for cost-effective visualisation of temporal events in dynamic scenes remain underexplored, particularly in terms of ensuring high usability and accessibility for individuals with limited animation skills.

To address this problem, we collaborated with forensic investigation experts to examine how immersive technology might democratise \CSR, making it accessible to those without specialised 3D modelling or animation skills. This approach could help investigators, legal professionals, and other stakeholders to more effectively visualise, analyse, and present crime scenes. Our research also evaluates whether such reconstructions enhance investigation, presentation, and perception, or whether they introduce challenges such as misinterpretation, bias, and further technical complexity. In summary, our contribution is fourfold.

\begin{itemize}
    \item A report detailing our co-design process with forensic/crime investigation experts highlighting their feedback and our insights regarding requirements for \XR \CSR (\autoref{sec:method}).
    \item A framework and implementation of a toolkit for rapid prototyping of animated crime scenes, see \autoref{fig:teaser} and \autoref{sec:system-design}, which we dub ``Criminator'', short for ``crime scene animator''.
    \item A study (\autoref{sec:userstudy})  involving two participant groups: laypeople (representing jury) and individuals trained in criminology (experts), reveals results (\autoref{sec:results}) concerning the usability and the potential for hypothesis testing, presentation, and training.
    \item A discussion of the broader implications and design recommendations for the HCI community within legal and visualisation contexts (\autoref{sec:discussion}).

\end{itemize}
\section{Background} \label{sec:related-work}
Crime scene investigations can vary greatly in complexity in terms of the evidence that must be collected and considered in analysis. Forensic examinations can employ a wide range of methods, such as 3D modelling and visualisation \cite{maneli2022csrsurvey,Balbudhe2025CSRreview}, DNA identification \cite{Butler2015-ij}, laser-based measurement \cite{Kudonu2023lasermeasure}, and statistical analysis \cite{lucy2013introstatforensic}. In this section, we focus specifically on visualisation and interaction methods that support human visual cognition for understanding, analysis, and presentation in judicial processes, excluding techniques that---while useful in forensic and legal contexts---do not directly serve these purposes. Within this scope, we first outline visualisation techniques and their respective strengths and weaknesses concerning legal criteria, then review \XR and animation capabilities for \CSR in greater depth, which are the primary focuses of this work.

\subsection{Benefits and Concerns of Visual Fidelity-Levels of CSR}

Despite technological advances in visual analytics, traditional methods such as sketching, photography and videography remain widely used in a crime scene investigation to quickly collect data and document observations due to the need for timeliness~\cite{maneli2022csrsurvey}. Depending on the nature of a criminal event, other visual-based methods may also be employed that provide more precise capture of crime scene details and therefore higher-fidelity reconstruction. However, it is crucial to understand how different fidelity-levels and technologies can be beneficial in a legal context, and at the same time what extra challenges they can impose.

Aforementioned traditional methods offer only 2D views of a crime scene. Sketches can be used to quickly map geometry and mark key details for communication, but they lack accuracy, are subject to human-error, are easily altered, and therefore are not reliable enough for court on their own. Photos and videos capture more detail and can verify sketches and may be less prone to manipulation and errors, yet they still flatten a 3D scene into 2D views at the time of capturing. The disconnected views and lack of navigable viewpoints increase cognitive load when trying to understand spatial relationships between items.

Computer graphics techniques that simplify 3D reconstruction have been used in crime scene analysis for decades~\cite{breuninger1995crime,murta1998modelling}. More recent techniques create high-fidelity, realistic 3D scenes through laser scanning~\cite{Raneri2018Laserscan}, video walk-through~\cite{Fei2024gaussiansurvey}, and photogrammetry~\cite{Villa2019photogrammetry}. Each of these techniques has advantages and disadvantages in different conditions~\cite{Dixit2019photogram3Dscan,Cunha2022laserphotocomp}. The need for manual modelling also remains for non-existing and damaged objects. Compared to photo and video evidence, these 3D model representations provide connected views and viewpoint movement, enhancing human comprehension and decision-making~\cite{horvath2024reconstruction}. 

Despite the benefits of 3D visualisations, creating such models requires equipment and skills that are not widely accessible in different jurisdictions. In addition, the high-fidelity and realism of visualisation techniques---regardless of whether they are reconstructed from accurate evidence or based on inference---have been found to be persuasive and could potentially deceive other involved parties, especially jurors \cite{schofield2009animating}. For this reason, Bailenson \etal ~\cite{BAILENSON2006} suggests that the more realistic visualisations and simulations are, the more interactive tools are necessary. Such interactive tools must enable creation and visual comparison of alternative hypotheses. 
In other words, the more advanced technologies for reconstruction become, the more important it is to provide accessible tools that ``open the black box'', supporting explorability and explainability of the provenance of the reconstructed crime scene model.
\XR is receiving increasing attention as a platform for 3D visualisations that enable more natural and simplified navigation and interfaces, despite the initial costs of hardware and software development. This area is discussed in detail in \autoref{sec:RW-CSRVR}.

Another important criterion for using and developing new techniques in a legal context is validation. Validating new techniques and methods from first principles can be a lengthy process. However, such techniques are often not entirely novel but instead represent variations of established methods. Recently published ISO 21043 Forensic Sciences permits variations and deviations from verified and validated methods to be used in a forensic examination as long as they are technically justified and recorded~\cite{morrison2025forensicscienceISO}. The standard also allows ``professional opinions'' and subjective interpretation for determining likelihood of a proposition using an interpretative method when it cannot be determined statistically. Although this does not grant a permission to a new technique to be deployed in legal settings directly, this opens a pathway to it.

Due to high dependency of judicial processes on human decision-making, a proper justification of a method should consider many human factors \cite{lee2022hfjurors,SMIT2018misleadevidence,kaye2009falsepersuasive}, including but not limited to, ease of perception, ease of use, human errors, bias, persuasiveness, fairness, and evidentiary/hypothetical distinctness. The many dimensions of human factors emphasise the role of HCI researchers to closely work with forensic professionals---for their professional judgement---to design interactive tools that technically justifiable for specific use cases with clear boundaries of intended use.

This perspective underpins our work: we develop a framework and a variation of existing tools---\CSR using \XR---through an expert-in-the-loop design process, and evaluate it with both criminologists and lay participants to determine whether the level of fidelity afforded by our tool enhances accessibility, supports perception and informed decision-making, reduces misjudgement, and facilitates hypothesis testing, or can also have negative effects on these aspects.

\subsection{CSR using XR}\label{sec:RW-CSRVR}
Due to the 3D nature of crime scenes, \XR technologies offer potential as analytical platforms.
\AR headsets~\cite{poelman_as_2012} and tablets~\cite{streefkerk_art_2013,tolstolutsky_experience_2021} have been used to annotate crime scenes and document them through photos and videos, accelerating the capturing process via virtual markers for indoor use cases. Gee \etal~\cite{gee_augmented_2010} introduced a system that enables remote experts to collaborate with on-site operators for large-scale mapping and data capture using handheld AR for indoor and outdoor areas. Similarly, Datcu \etal~\cite{datcu_handheld_2016} examined handheld \AR combined with shoulder-mounted cameras to provide collaborative, distributed guidance for on-site operators from remote experts. These methods improve efficiency and comprehension of on-site investigation.

While \AR has been utilised for collecting and analysing data at crime scenes, \VR has mostly been applied for off-site examination in previous studies. For example, S\"uncksen \etal~\cite{suncksen_preparing_2019} used \VR to visualise a 3D-modelled scene with numerous characters and bullet trajectories to support detailed inspection. Due to advances in 3D scanning, Reichherzer \etal~\cite{reichherzer_bringing_2021} created point clouds from 3D scanner data, enabling investigators to explore and analyse a digital twin of a crime scene in \VR. In another related study, Rinaldi \etal~\cite{rinaldi_crime_2022} combined a scanned crime scene in \VR with photographs placed at their real position within the virtual scene, providing the off-site investigation  with spatial context for photographs. Our work builds on these studies,   visualising spatial context in \VR. Notably, however, we employed only very recently available Gaussian Splatting techniques~\cite{kerbl2023gaussian} to render captured physical environments in significantly higher fidelity.

\XR offers trainees an accessible and cost-effective platform to repeatedly practise crime scene investigation protocols in a realistic context, experiencing the complexity and diversity of physical crime scenes first hand  ~\cite{albeedan_designing_2024,albeedan_evaluating_2023,Chang28062024}. Gamification has enabled the creation of more interactive environments by simulating crime scenes, collecting clues, and conducting examinations in \VR~\cite{Woodward2021,vizitech_vr_forensics,victoryxr_csi_forensics}.

In addition to static representations, forensic animations have been used to analyse traffic accidents~\cite{BUCK2007dataanim}, airplane crashes~\cite{ctx2264014146750001751}, bullet trajectory analysis~\cite{Galligan2017trajectory}, and human posture simulation~\cite{Villa2017humanposeanim}  for both crime/incident investigation and courtroom presentation. However, scientific discussion and research particularly regarding effective and accessible tools remain notably limited.

To conclude, research has found benefits in 3D scene reconstruction and using \XR for inspection, training, and presentation. However, past work has focused on static reconstruction, animation of very specific scenarios, or gamification. The potential of \XR  for reconstructing temporal relationships between objects in realistic context is underexplored.

\subsection{Authoring Animation using XR}

While 3D animation is widely used in \XR applications, it typically requires advanced desktop software, skilled 3D modellers and animators, specialised equipment, significant budget and time investment. However, in recent years, attempts have been made to create animations within \XR environments to leveraging 3D interaction, as well as direct and embodied manipulation.

Past studies focused on non-humanoid~\cite{Cannavo2019vrcharanim} and humanoid~\cite{Zhou2024timetunnel,garcia2019spatial,chen2023doubledoodles} character animations with low-level control of joint positioning and key-frame recording. This approach---inspired by desktop equivalents---has been adopted by other studies as well. Vogel \etal~\cite{Vogel2018animvr} developed a \VR tool with extensive capabilities for creating animations, allowing users to directly manipulate and alter models and animations. Similarly, Zhang \etal~\cite{Zhang2020Flowmatic} explored how a visual programming interface, similar to Unreal Engine's Blueprints, can support the creation of interactive scenes in \VR. However, both remain relatively complex to use, requiring skill and time to create an animation. 
By contrast, collaborative authoring~\cite{Pan2020PoseMMR} as well as artistic and cartoonish animation~\cite{Quilez2016Quil,Dario2018vranim} are among the efforts to ease direct animation creation in \XR.

While these tools advance animation creation, a faster solution is necessary for frequent workflows. Li \etal~\cite{li2024anicraft} developed an AR-based tool that enables rapid animation prototyping with physical proxies, reducing required skill but facing challenges in proxy construction and managing physically-constrained object interactions. Similarly, Wu \etal~\cite{Wu2023ImpersonatAR} provides an AR tool for capturing and viewing character animations within a scene, and only a limited flexibility in terms of interaction---such as triggering a button---with other virtual objects or the environment, making it mainly suitable for developing step-by-step training materials and quizzes.

Among \XR animation authoring tools, to the best of our knowledge, Flipside XR~\cite{flipsidexrMetaverseSocial} is the most capable of meeting our requirements for animated crime scene reconstruction. It is a feature-rich \VR application designed  for fast 3D \VR social media content creation. It enables the rapid creation of animations with relatively low detail, and animating characters from first person view. It also allows for the expansion of objects and dynamics for developers. However, it is a closed-source, proprietary software application. While it supports third-party plug-in development, the current version lacks the flexibility required for developing customised features for crime scene reconstruction.

\section{Method}\label{sec:method}
\begin{figure}
    \centering
    \includegraphics[width=\columnwidth]{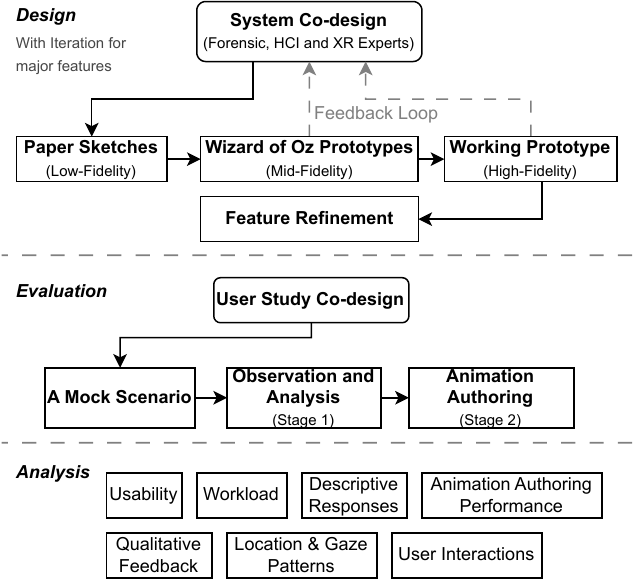}
    \caption{Our methodology for design, evaluation and analysis.}
    \label{fig:codesign-methodology}
    \Description{This flowchart is divided into three sections: Design, Evaluation, and Analysis. Design section: The process begins with System Co-design, which leads to Paper Sketches, then to Wizard of Oz Prototypes, and finally to a Working Prototype. This sequence represents progression from low to high fidelity. From the Working Prototype, the process ends with Feature Refinement. There are two feedback loops: one from Wizard of Oz Prototypes back to System Co-design, and another from Working Prototype back to System Co-design. Evaluation section: This section begins with User Study Co-design. From there, it branches to Mock Scenario Design, which is followed by two sequential stages: Observation and Analysis (first stage) and Animation Authoring (second stage). Analysis section: This section consists of seven unconnected components: Usability, Workload, Descriptive Responses, Animation Authoring Performance, Qualitative Feedback, Location and Gaze Patterns, and User Interactions.}
\end{figure}

We conducted a co-design process \cite{Sanders2008Cocreation} with forensic investigation experts as outlined in \autoref{fig:codesign-methodology}. Co-design is a prevalent methodology in \HCI, specifically \XR studies \cite{Bryant2024codesign,Acton2024codesign,Shen2023dementia}, to build usable tools and prototypes for the target(s) groups including forensic applications \cite{pooryousef2023working,Lantta2024violenceprevent}.  In total, we incorporated the perspectives of six forensic experts both prior to and during the design process (see \autoref{tab:experts_details}). Expert 1 and 2, who maintained continuous involvement, later joined the authors of this paper to contribute in validating user study's results, writing and improving comprehension of the paper, as it is also seen in highly specialised domains (\eg \cite{besancon:hal-01795744,Besancon:2020:RAR,pooryousef2023working}).
 
% Co-design diagram

As a result, the development team (the authors of this paper) comprises specialists in forensic science, \HCI and \XR. To facilitate more effective communication among experts from different fields, we employed multiple levels of prototyping:

\begin{enumerate}[leftmargin=*,itemsep=1pt, topsep=2pt, parsep=0pt, partopsep=0pt]
    \item First, ideas were sketched on paper (low-fi prototyping).
    \item  Once the concepts became more refined, we followed a Wizard of Oz approach \cite{green1985wizard}---here, creating a set of demonstrative user interfaces and captured animation using a motion tracker system to better visualise the overall structure and capabilities of the toolkit in communication with the forensic investigation experts.
    \item Finally, a high-fidelity prototype was developed when a consensus was reached on including a given feature. However, all features were not implemented at once; rather, for each new set of features, these phases were more or less repeated.
\end{enumerate}

We then evaluate the usability and effectiveness of our developed prototype through a user study consisting of two stages for observation and animation authoring capabilities, \autoref{sec:userstudy}.  It will gauge how usable, effective and limited this toolkit is for observation and animation authoring. More details about the questionnaires and test are provided in \autoref{sec:results}.

\begin{table}
    \centering
    \caption{Experts' details involved in the co-design process.}
    \Description{This table has four columns: row number, co-author, profession, and years of experience. It contains six rows of data, excluding the header row, with one row for each expert participant.}
    \begin{tabular}{|c|c|c|c|}
        \hline
        \textbf{\#} & \textbf{Co-author?} & \textbf{Profession}                                                     & \textbf{\begin{tabular}[c]{@{}c@{}}Years of\\ Experience\end{tabular}} \\ \hline
        1           & Yes (4th)           & \begin{tabular}[c]{@{}c@{}}Crime Scene\\Reconstructionist\end{tabular} & 10+                                                                    \\ \hline
        2           & Yes (5th)           & Forensic Scientist                                                      & 50+                                                                    \\ \hline
        3           & Yes (6th)           & Forensic Odontologist                                                    & 20+                                                                    \\ \hline
        4           & No                  & Crime Scene Examiner                                                       & 30+                                                                    \\ \hline
        5           & No                  & \begin{tabular}[c]{@{}c@{}}Forensic Scientist with\\Law-Enforcement Agency\end{tabular}                                                      & 10+                                                                    \\ \hline
        6           & No                  & \begin{tabular}[c]{@{}c@{}}National-level Police Officer\\and Forensics Researcher\end{tabular}                                                    & 10+                                                                    \\ \hline
        \end{tabular}
    \label{tab:experts_details}
\end{table}
\section{Criminator System Design}\label{sec:system-design}
To establish a coherent design rationale for our system, we gathered requirements and criteria from discussions with forensic experts.

\subsection{Experts' Insights}
The 2023 International Association of Forensic Sciences (IAFS) meeting \cite{ROSS2025whereto} emphasised the necessity of a scientific approach and adoption of new technologies in forensic investigations. From this perspective, the scientific approach includes observation, critical thinking, problem-solving, hypothesis formulation, and reconstruction. In addition, through discussions with experts in this field and informed by OSAC guidelines \cite{OSAC2025-N-0004}, we have concluded that the technology for \CSR must support the following principles:

\begin{itemize}
    \item The \CSR methodology requires the application of critical thinking and the scientific method.
    \item \CSR should not rely on logical flow, common sense, personal beliefs or expected behaviours to define the sequence events, rather it must be supported by physical evidence, scene context and/or physical laws.
    \item A reconstruction requires contextual information to guide the analysis. However, reconstructionists must take steps to mitigate impacts of cognitive biases on their work.
\end{itemize}

\noindent Such system must also meet criteria to ensure practical applications:

\begin{description}
    \item[Usability] Once completed and validated, it should be accessible and readily usable for any crime scene investigator involved in the investigation of complex crimes with minimal training required.
    \item[Timely] The tool should be capable of displaying different, viable scenarios in a short space of time.
    \item[Flexibility] It requires flexibility to adapt to the changing circumstances and information related to crime investigation and the impact these have on reconstructing the crime.
    \item[Information transfer] It should be easily demonstrated and explainable to decision makers (\eg the judiciary and juries) who are involved in the justice process, especially where viable alternatives are demonstrated. Courts are adopting an increasingly favourable attitude to the supportive use of technology in the presentation of expert evidence \cite{schofield2016cglegal}. Once a crime scene reconstructionist has determined the most likely scenario(s), the crime scene prototype(s) could be demonstrated to the court, judge and jury, to assist them in their decision-making process.
\end{description}

\subsection{Mechanisms}
\label{sec:mechanisms}
\begin{figure}
    \centering
    \includegraphics[width=\columnwidth]{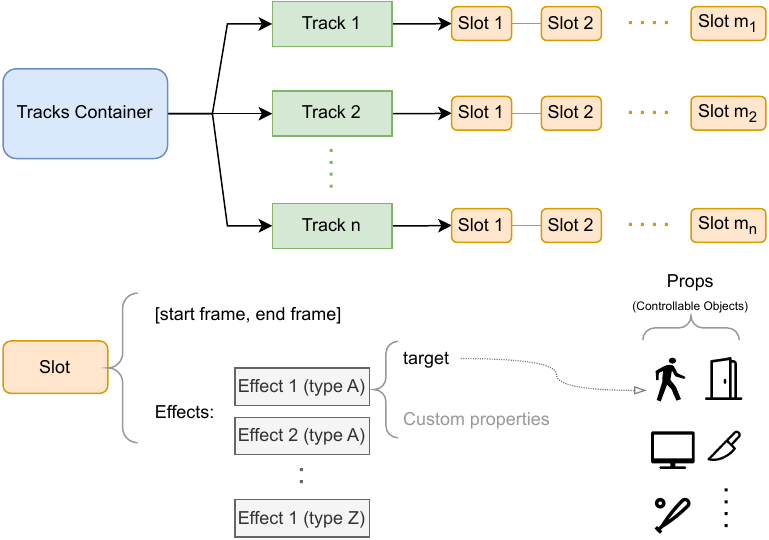}
    \caption{Overview of \CSR in \VR mechanism.}
    \Description{The diagram shows the mechanism of the animation authoring tool. At the top, a track container holds n tracks. Each track contains an arbitrary number of slots. At the bottom, the diagram provides details of a single slot. A slot is defined by its start frame and end frame, and it contains a list of attached effects. Each effect item has a type, and all types of effects include at least a target parameter plus additional custom properties specific to that effect. Targets can reference props or controllable objects in the scene such as a human avatar, a door, a computer, or a knife.}
    \label{fig:csivr-mechanism}
\end{figure}

To implement a prototype system for reconstructing animated crime scenes, we first developed a generic mechanism or framework inspired by regular video editing tools. \autoref{fig:csivr-mechanism} illustrates an overview of the underlying mechanism behind the system. In this figure there is a track container that holds multiple tracks. Each track can have multiple time slots. Each slot has a start frame, end frame, and a set of effects. Effect is an abstract entity that can be developed for a wide range of effects. It includes a \textit{target} property that specifies which object the current effect is applied to. Each implemented effect may have additional parameters as needed. Additionally, each implemented effect should store its animated data and retrieve it as necessary. Controllable object is also an abstract entity that represents props and objects that can receive effects, facilitating the extension of both effects and props (\ie controllable objects) with minimal effort. For this study, we implemented several types of controllable objects and effects, detailed in \autoref{sec:system-features}.

There are also some constraints in this architecture that---while not explicitly visible---enhance the system flexibility and prevent excessive complexity in managing animations. Here are the key constraints:
\begin{enumerate}
    \item Between two tracks the one that has a lower number (\ie the upper track) has higher priority. This means  that if two effects control the same attribute of the target object, the effect with higher priority overrides any changes applied in that frame.
    \item Multiple instances of the same type of effect may be present in a single slot, only if they have different target objects.
    \item Slots within the same track do not overlap. This simplifies priority management already determined by track priority constraint (Constraint 1).
\end{enumerate}

\subsection{Criminator Features}\label{sec:system-features}
The described mechanism requires a user interface to apply the structure and constraints, implementation of the necessary effects and props for crime scene investigation, and an animation recorder and player to run the created tracks, slots, and effects.\footnote{Source code is provided at \url{https://github.com/vahpy/XR-Crime-Animator}}

\subsubsection{User Interface}
\begin{figure}
    \centering
    \includegraphics[width=\linewidth]{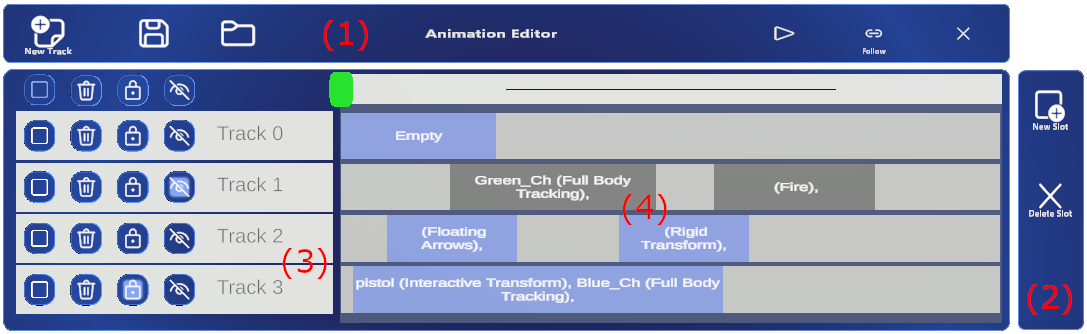}
    \caption{Animation Editor \UI.}
    \Description{This figure shows the animation editor interface, which is divided into four main areas: 1) Top bar: contains buttons on the left for creating, saving, and loading tracks, and on the right for playing the animation or closing the editor. 2) Right bar: provides two buttons, one for creating a slot and one for deleting a slot. 3) Track control area: displays all tracks, each with control buttons to hide, lock, delete, or select the track. 4) Slot control area: within each track, slots are shown with their length, applied effects, and target properties.}
    \label{fig:user-interface}
\end{figure}

The main \UI in this toolkit is the animation editor panel. This panel provides necessary functions needed to manage tracks, slots, effects and the parameters. The main functions accessible on this panel (numbered as per \autoref{fig:user-interface}) are:

\begin{description}
    \item[1) Top bar] creating a new track, storing and loading animation, playing and pausing animation, following mode, and hiding the panel.
    \item[2) Right bar] creating and deleting a slot.
    \item[3) Track control area] sorting tracks, disabling or making their effects invisible, locking modification, and deleting them.
    \item[4) Slot control area] time slider (green button), drag and dropping slot in a different location in the same or different track, as well as trimming a slot duration by dragging each end of the slot.
\end{description}

\begin{figure}
    \centering
 \includegraphics[width=0.68\linewidth]{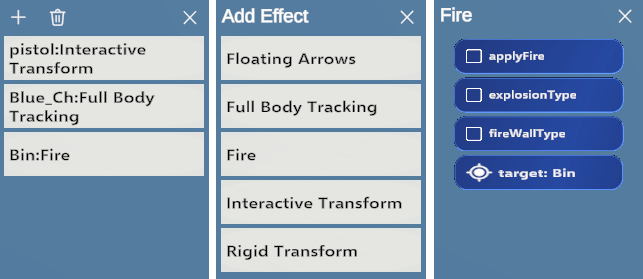}
    \hfill
\includegraphics[width=0.28\linewidth]{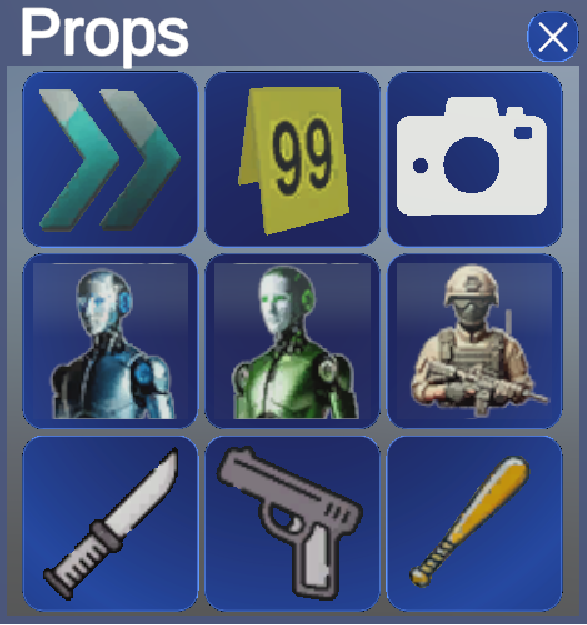}
   % \vspace{-3mm}
    \caption{Effects (3 left) and props (right) panels. The Effects section includes: (left) a list of current effects attached to a slot, shown as ``target prop:effect''; (middle) a list of all available effects that can be added to a slot; and (right) the parameters of the selected effect.}
    \Description{This figure contains four panels arranged from left to right. The first panel shows three effect items, each with a target prop and its associated effect. The second panel, titled "Add Effect", displays a list of available effects such as "floating arrows", "Full body tracking" and "Rigid Transform". The third panel, titled "Fire", presents the parameters of the Fire effect. These include options such as Apply Fire, Explosion Type, Firewall Type, and a Target set to "Bin". The fourth panel, titled "Props", displays nine prop items. These include crime scene–related objects such as a camera, two robotic avatars, one police avatar, a knife, a gun, and other props.}
    \label{fig:editor-panels}
\end{figure}

Upon selecting a slot on the editor, a small rectangle panel is shown next to the editor, and a user can add and manage effects. As seen in  \autoref{fig:editor-panels}, the effects panel has three states:
\begin{description}
    \item[Slot's effects] shows a list of attached effects to the slot, with buttons for adding an effect or deleting one.
    \item[Effects list] shows a list of available effects to be added to the slot.
    \item[Effect parameters control] provides control over parameters of an effect.
\end{description}

As seen in \autoref{fig:editor-panels}, users can add a prop to the scene from the \textbf{Props Menu}. This list can be expanded to include as many props as necessary.

In addition to regular selection, manipulation, and navigation using the controllers, users can also start and stop recording, toggle the animation editor and props menu, and switch between ground-level and top-down views directly via the controllers (see the guiding labels in \autoref{fig:user-interface}). They can also hold grip buttons to zoom in and out along the depth axis by adjusting the controller's distance, and to move (pan) the view by shifting the controller in any direction, which is especially useful in a top-down view for ``birds-eye'' navigation.

\subsubsection{Effects and Controllable Objects for \CSR}
As mentioned \autoref{sec:mechanisms}, effects and controllable Objects are two abstract entities in this architecture that can be implemented as add-ons depending on the application. To better handle the diversity of objects, we also implemented two levels of abstraction such as \textit{triggerable} and \textit{non-triggerable} controllable objects. In this prototype, we extended the toolkit to support a typical crime scene scenario such as the one described in \autoref{sec:scenario}. The list of controllable objects can include human characters, criminal instruments, pieces of furniture, buildings, investigation markers, notes, measurements, and photos. Most of these objects are quite straightforward to implement. However, it is worth explaining the environment (such as buildings and furniture) implementation in more detail. 

While it is possible to create a building with 3D models, we also utilised Gaussian Splatting \cite{kerbl2023gaussian}---a novel and high-quality 3D scanning and volume rendering technique---for realistically representing the actual crime-scene environment. With this approach, a significant portion of modelling the environment and context through 3D models is simplified by a quick scan with a phone, which also provides a more realistic context. While in the past, 3D capture of real-world environments relied on expensive equipment such as tripod-mounted LIDAR, phone-based scanning is becoming ubiquitous, being easier to use, faster (hours vs. minutes), cost-effective, and more accessible for defence and police departments across the world, and, as such easier to integrate in workflows. Previous research \cite{Stevenson2024PhoneScanForensic} also shows that phone 3D scanning offers acceptable results for bloodstain pattern analysis in forensic examinations, which meets the quality requirements of our study.

Effects are more complex and their implementation difficulty highly depends on the type of effect. They can modify the state and properties of objects or even create new objects. Additionally, they can access various features of the Unity game engine, such as particle systems, audio, and physics. The key implemented effects in this toolkit include:

\begin{description}
    \item[Full Body Tracking] A crucial effect for character animation, utilising the Meta Movement SDK to estimate full-body joint positions for natural movement. This effect provides estimations rather than precise position tracking.
    \item[Rigid Transform] Captures changes in an object's transform properties (position, rotation, and scale) per frame or at keyframes, interpolates between them for smooth transitions, and applies physics such as collision and gravity configured by the user.
    \item[Interactive Transform] Records the state changes of interactive props, such as weapons or other triggerable props.
\end{description}

While not directly used in the study, we also implemented a ``Floating arrows'' (animated arrows showing connections) and a ``Fire'' effect, with their parameters such as the type of fire as shown in \autoref{fig:editor-panels}. Except for the two latter effects, the other effects store animation in the Unity standard animation format.

\subsubsection{Animation Recorder and Player}

\textbf{Playback.} On playback start, an effect receives a \texttt{START} signal if:
\begin{itemize}[leftmargin=*,itemsep=1pt, topsep=2pt, parsep=0pt, partopsep=0pt]
    \item The user presses play with the time slider in the slot's range.
    \item The time slider reaches the slot's start frame.
\end{itemize}

During playback, the effect receives an \texttt{UPDATE} signal per frame; If paused or reached the effect's end, it receives a \texttt{PAUSE}.

While the \texttt{UPDATE} signal eliminates the need for a \texttt{START} signal for many types of effects, distinguishing them simplifies implementation, especially when continuous playback is handled differently from per-frame updates or one frame playback.

Unlike most video players, on the playback start, our proposed animation player requires a quick playback of all effects in all tracks up to the current frame, even if the current frame has already passed those effects' end frame. However, the implementation may vary depending on the effect, as some effects always have their final state at the end frame.  To optimise storage, effects may record only changes from the previous frame, necessitating quick playback to determine the final state. Obviously, effects must store the initial state of the target object at the first frame.

\noindent\textbf{Recording.} 
To record an effect, whether continuous or at specific key frames, animation recorder notifies only the currently selected effect to store changes to its target object(s). All other effects switch to playback mode instead. Therefore, a user see the animation in \VR, while recording changes to an effect.

For simplicity, the recorder also appends frequently-used effect, such as object movement effect, automatically to the selected slot when needed. This reduces steps needed for creating animations involving multiple objects and effects. For instance, when animating a character and it picks up a prop or shoots with a gun.
\section{User Study}\label{sec:userstudy}

With the guidance of forensics experts (\autoref{tab:experts_details}), we designed a user study based on a task analysing a mock crime scene. The context is a dynamic scene with an ambiguous sequence of events, multiple individuals, and areas where viewing angles could either enhance or obstruct the observer's view.
\subsection{Scenario}\label{sec:scenario}
\begin{figure}
    \centering
    \begin{subfigure}[b]{\columnwidth}
        \includegraphics[width=\columnwidth]{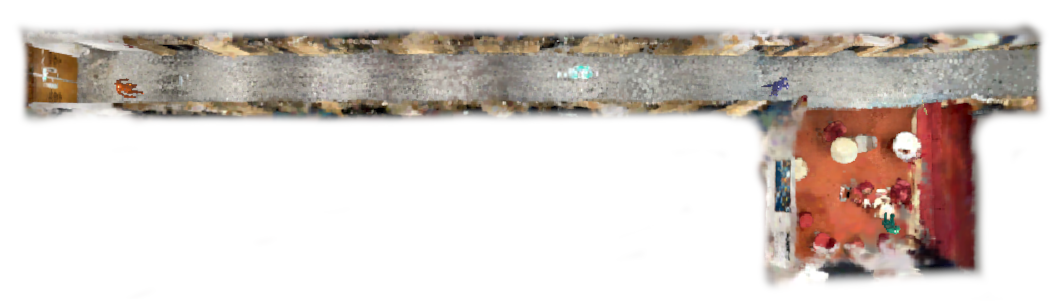}
        \caption{The scenario’s environment from top-down view.}
        \Description{}
        \label{fig:scene-map}
    \end{subfigure}
    \hfill
    \begin{subfigure}[b]{\columnwidth}
        \begin{tabular}{cc}
            {\includegraphics[width=0.45\linewidth]{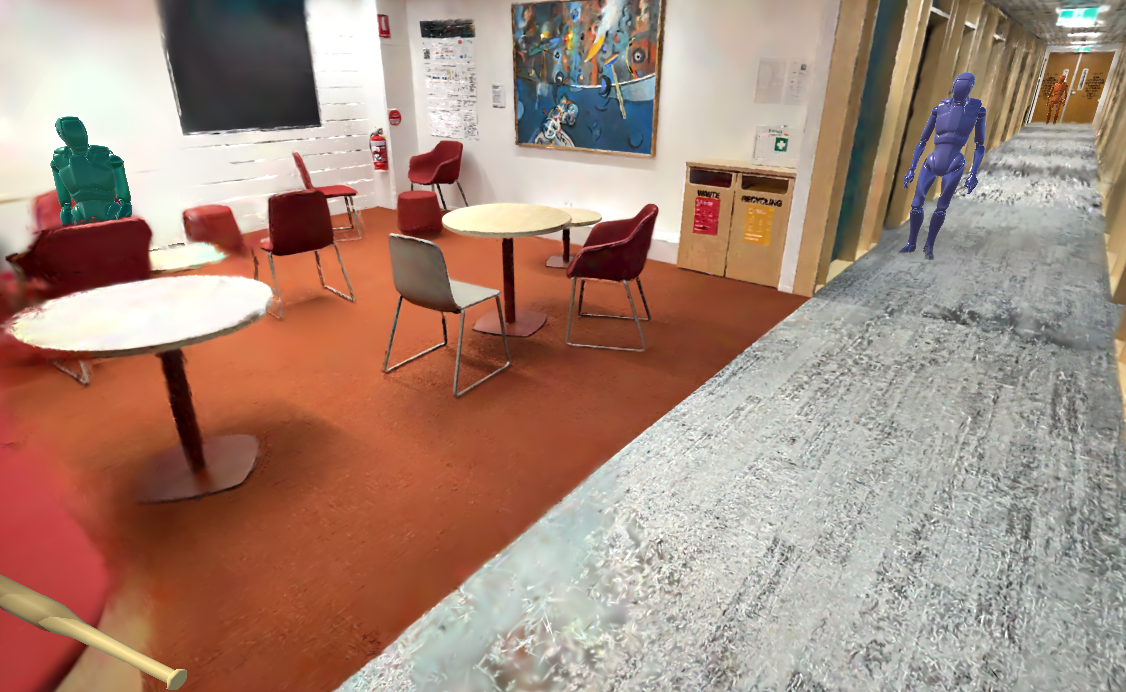}} &
            {\includegraphics[width=0.45\linewidth]{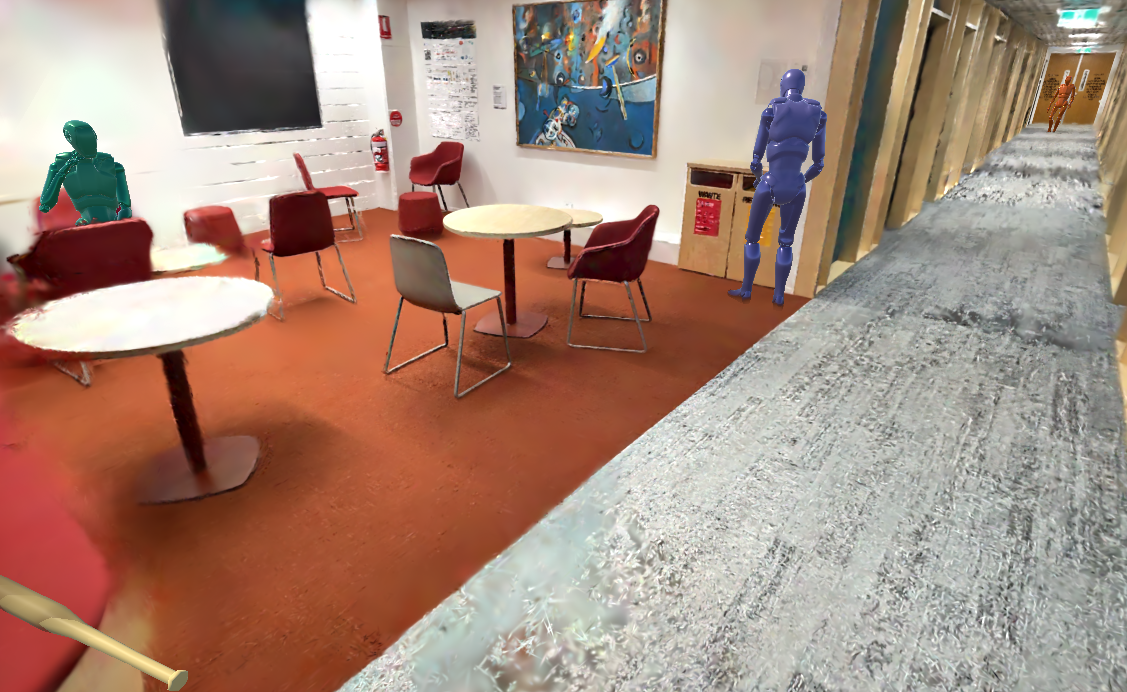}} \\
            
            {\includegraphics[width=0.45\linewidth]{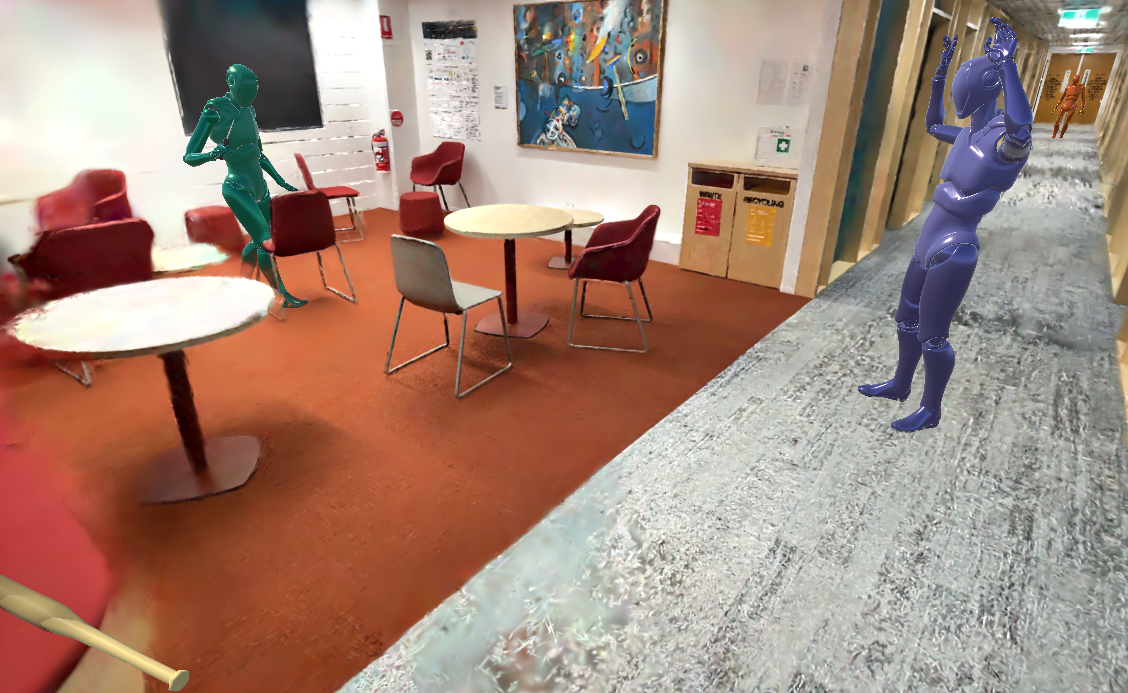}} &
            {\includegraphics[width=0.45\linewidth]{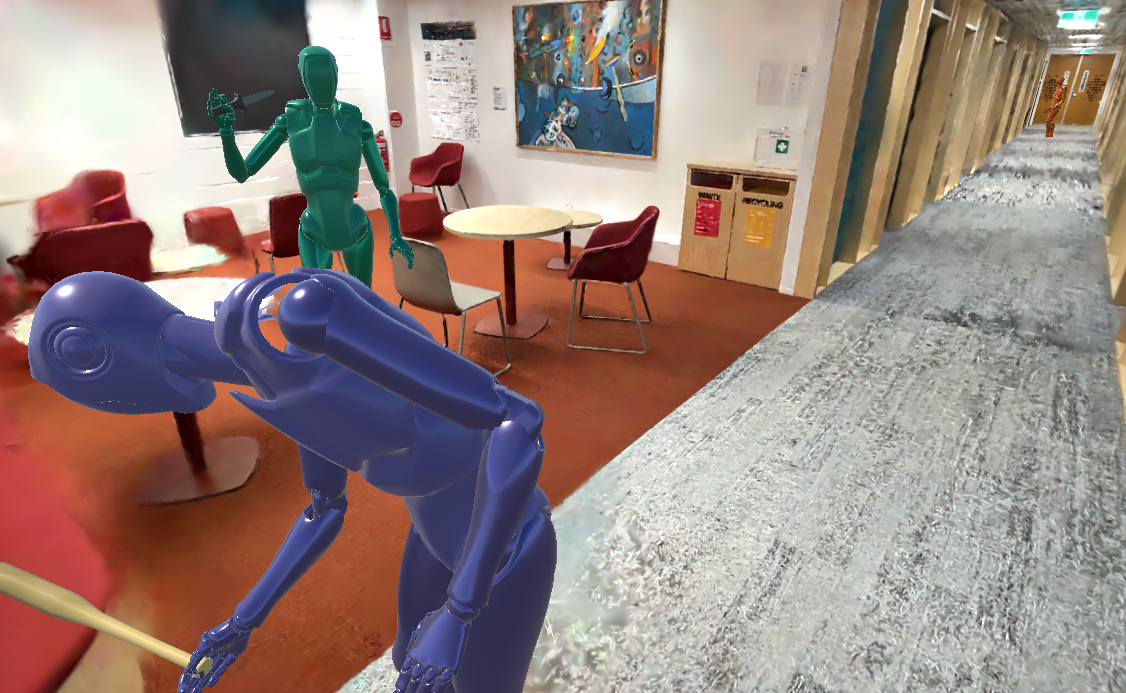}} \\
            
            {\includegraphics[width=0.45\linewidth]{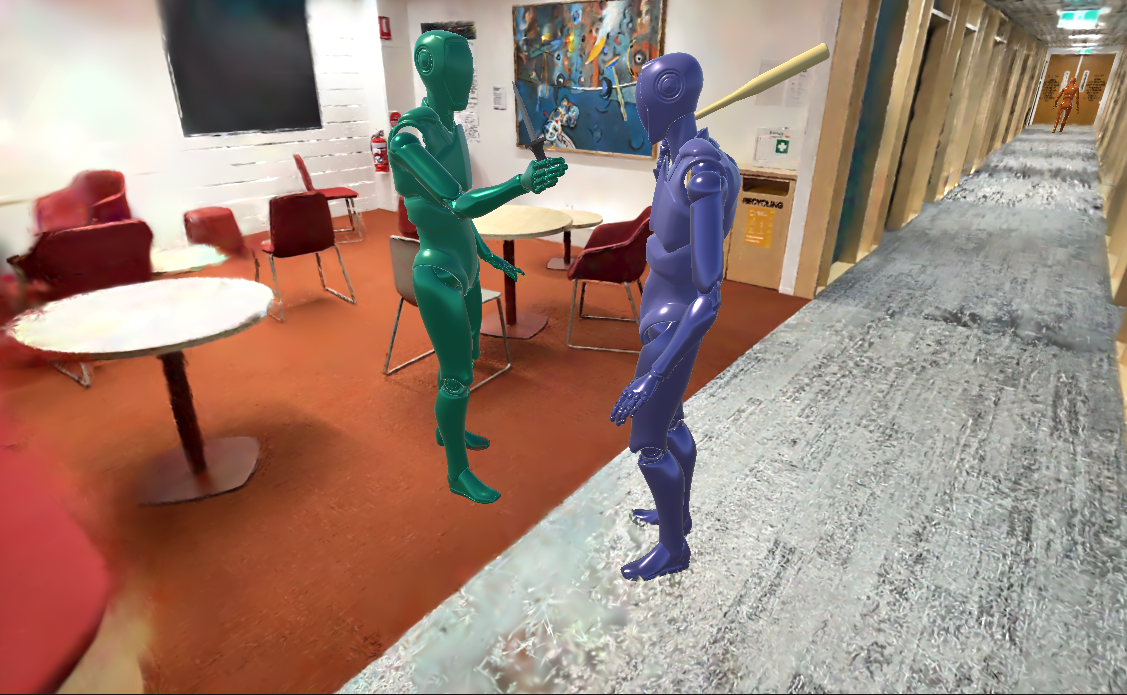}} &
            {\includegraphics[width=0.45\linewidth]{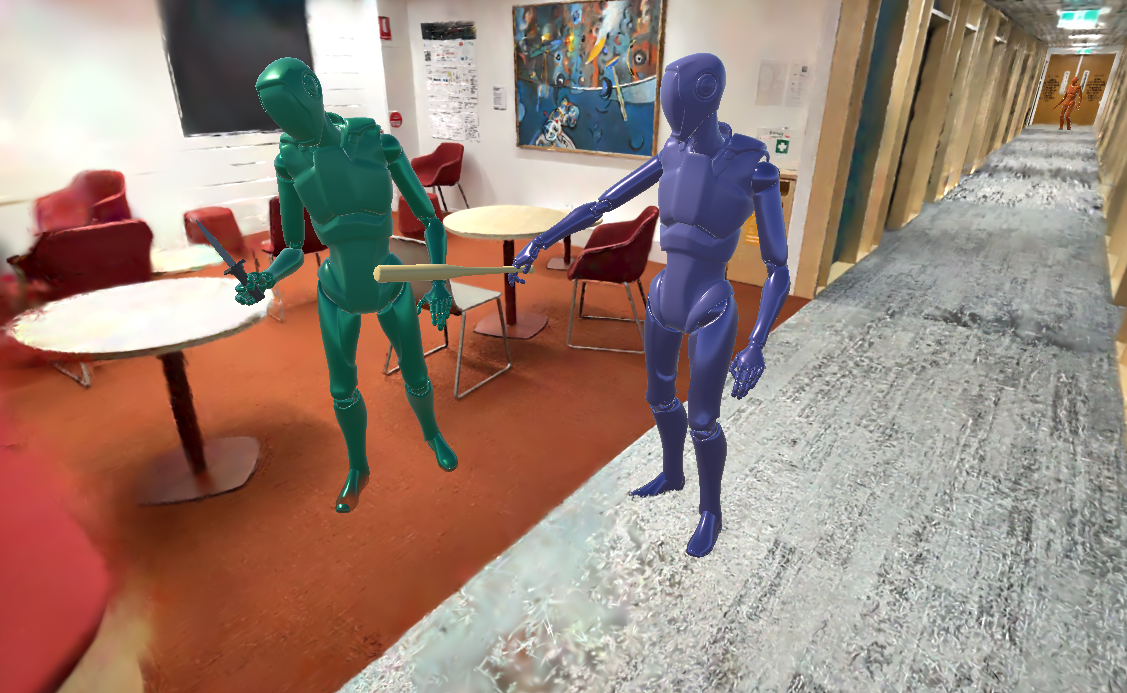}} \\
            
            {\includegraphics[width=0.45\linewidth]{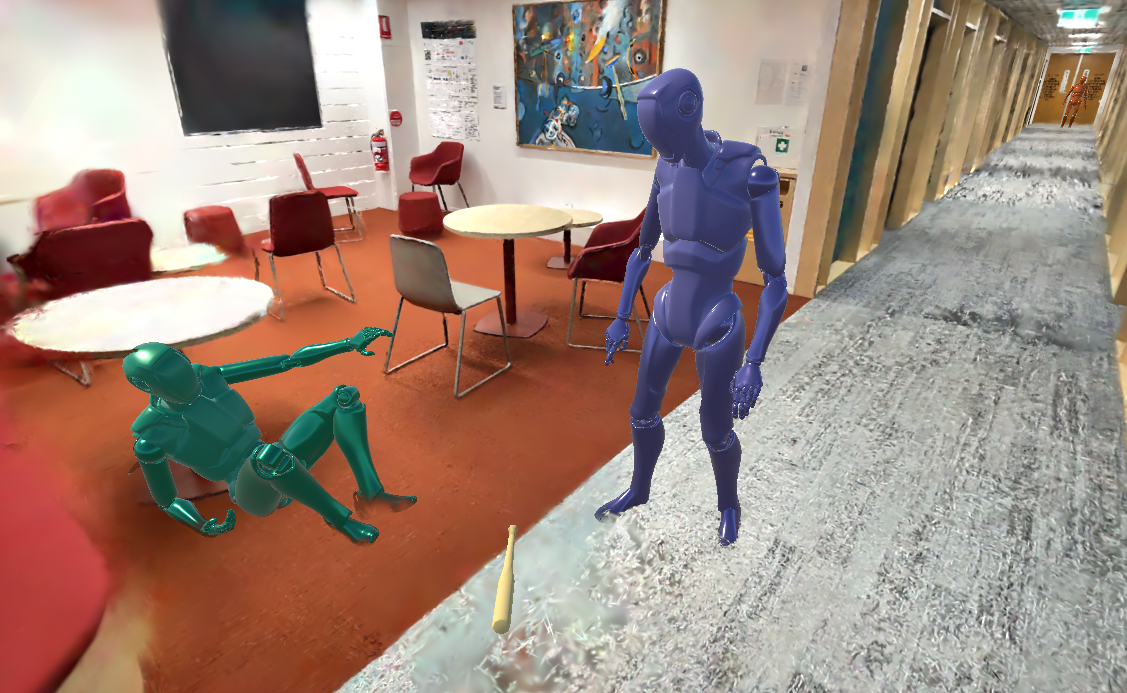}} &
            {\includegraphics[width=0.45\linewidth]{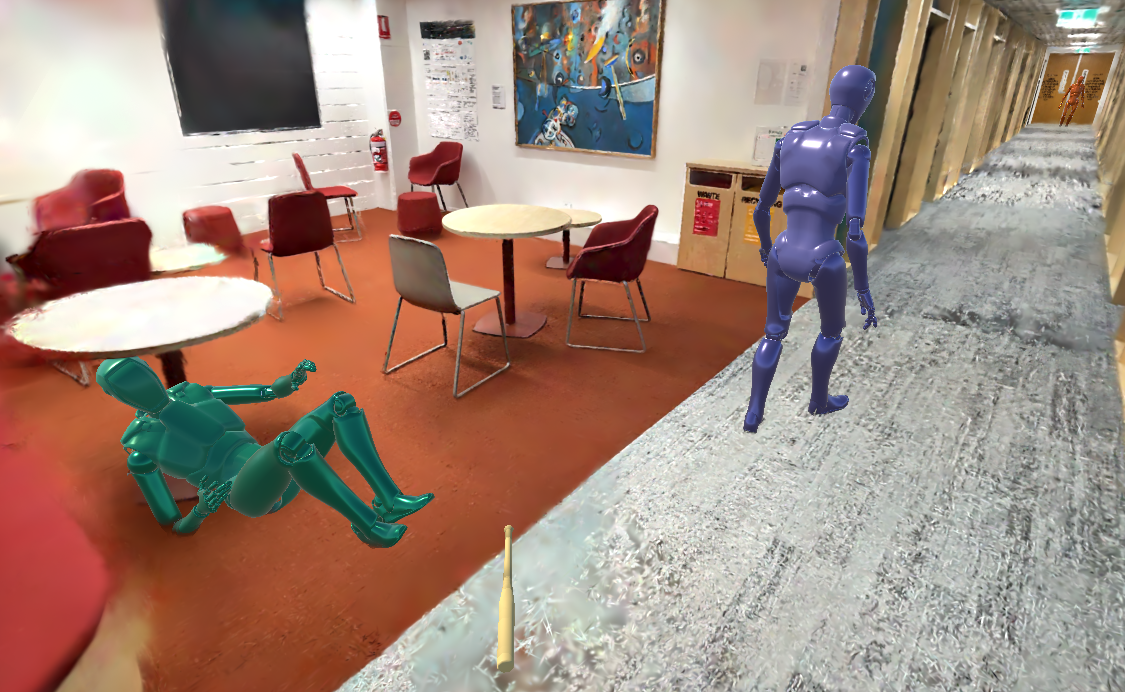}} \\
        \end{tabular}
        \caption{Key moments of the user study's scenario, chronologically ordered from left to right and top to bottom.}
        \Description{}
        \label{fig:scenario-steps}
    \end{subfigure}
    \caption{The overview of the user study's scenario.
    }
    \Description{This figure has two parts: (a) and (b).
(a) Top-down view of a long hallway with a small study area on the right. Two avatars are walking down the hallway, while one is sitting in the study area..
(b) Eight key frames of a crime scenario. In the sequence: a person approaches the study area, throws trash in a bin, is then threatened by a second person wielding a knife, grabs a bat and engages in a fight until the attacker falls, then drops the bat and walks away. In all frames, another person, as a witness, is visible at the end of the hallway.}
    \label{fig:scenario}
\end{figure}

A crime occurs in a building with a long hallway and a small lounge area, as seen in \autoref{fig:scenario}. The scenario involves three individuals: a witness (orange person), an attacker (green), and a defender (purple), as depicted in \autoref{fig:scenario-steps}. This figure illustrates key moments in the scenario. At the beginning, the witness is walking toward the study room from the end of the hallway, while the defender is heading toward the room to throw trash in the bin. As this happens, the attacker notices the situation and grabs a knife that was on a table. The defender raises both hands to signal that they mean no harm, but the attacker continues walking toward them. The defender then notices a baseball bat on the couch, grabs it for protection, and prepares to defend themselves. At this point, the witness senses something unusual and turns back toward the door. The attacker attempts to injure the defender with the knife. The defender begins defending themselves and strikes the attacker with the bat twice, knocking them to the ground. As the defender strikes, the witness looks back to see what is happening. Afterwards, the defender drops the bat, walks away, then after a few steps returns to check on the attacker.

\subsection{Tasks}
The study consists of two stages, approximately one hour total. The first stage involves determining whether the current tool provides an easy way to observe a crime scene, and how the viewing angle influences interpretation. This aspect has been investigated in previous studies, allowing us to evaluate whether our findings align with existing research. In the second stage, we focus on the animation creation task, which is the primary goal of our proposed tool. Since no comparable tool for authoring is available, comparing usability and task load scores between the two stages provides an indirect way to assess how our authoring tool perceived  relative to observation-only tools.

\subsubsection{Stage 1. Observation and Analysis}
In this stage, the participant is required only to observe the pre-generated animation of the scenario from different perspectives.
\begin{description}
    \item[Task 1] The participant is allowed to view the animation only from approximate areas where the witness is walking.
    \item[Task 2] The participant views the animation from a top-down perspective (at an approximate height of 20 metres) without restrictions on movement. Additionally, they may move up to 10 metres closer to the ground if necessary.
    \item[Task 3] The participant can observe the animation from inside the building without any restrictions.
\end{description}
After each task, participants are asked to answer the following questions:
\begin{itemize}
    \item \textit{What actions did the purple character take, and what was their intention?}
    \item \textit{What could the witness see, and how many individuals did they identify in the scene?}
\end{itemize}
Additionally, we ask about the participants' level of confidence in their responses after Task 1 and how they revise it after Task 3. Participants may replay the animation as many times as needed until they feel prepared to answer the questions.

\subsubsection{Stage 2. Animation Authoring}
In this stage, the participant must take on the role of the purple character and replicate the animation until they deem it correct. To accomplish this, they are required to set up an animation by adding a purple character and a baseball bat. In addition, they need to configure necessary settings for animation creation, including creating a track, a slot, adding body-tracking effects to the slot and assign the character as its target.

\subsection{Tests, Questionnaires and Rationale}\label{sec:test_rationale}
Participants are assessed individually and collectively based on their responses and performance in both qualitative questions and quantitative measures, as follows:

\noindent\textbf{Usability and workload.}
System Usability Scale (SUS) \cite{brooke1996sus} and NASA Task Load Index (TLX) \cite{HART1988nasatlx} are commonly used questionnaires in \HCI studies \cite{Kosch2023cognitiveloadsurvey} to measure usability and perceived workload of a system/tool/method. We used raw NASA-TLX, since the second part of the survey may be confusing for participants \cite{besancon:hal-01436206} and is also common in the field (\eg \cite{besancon:hal-01436206,sereno:hal-03699232}). While SUS and NASA-TLX scores provide valuable insights, these should typically be assessed against a baseline. This study focuses on animated crime scene authoring, however, as reviewed in \autoref{sec:RW-CSRVR} there is no similar tool in \XR to compare as a baseline. Desktop software capable of making character animations is overly complicated, requiring significant training, which is not feasible for non-experts in 3D modelling. 

\noindent\textbf{Descriptive Questions for Stage 1.}
Stage 1 explores benefits and challenges of crime scene observation and analysis in \VR from trained forensic experts and untrained people. This represents different stakeholders involved in a crime examination and judgement, through their qualitative responses to the aforementioned questions for Task 1 to 3. These questions have been designed to assess how observing animated crime scene from different angles affects an observer's perception of events and intentions.

\noindent\textbf{Animation quality assessment.}
We conduct a blind expert review process in two rounds (reviewers are the first and second experts in \autoref{tab:experts_details}). The animations are randomised for each reviewer and each round. Each reviewer watch each animation and rate it from 1 (poor quality) to 5 (excellent quality) with notes about the criteria they deducted a score. Given that the dataset consists of 17 animations featuring similar actions and that participant number and group are undisclosed, we anticipate that reviewers will be unlikely to recall their previous ratings for a specific animation.\footnote{Additional details on the assessment design provided in the Supplemental Material
.}

\noindent\textbf{Location and gaze patterns.}
We collect the participants' location and gaze direction in Stages 1 and 2 for each task separately. For Stage 1, it is useful to collectively analyse what locations participants usually choose to observe an animated scene in \VR from different angles and their respective field of view. For Stage 2, the participants' location can show how the path in the character animation could vary among participants, which complements the animation quality assessment.

\noindent\textbf{Observations from user interactions}
Some behavioural data remains hidden from the aforementioned tests and data collection, which also may not be included in the participants' reflection on their performance.
This not only provides insights to the \XR and forensic research communities in future research, it also helps in understanding the reasons behind some of the resulting data.

\subsection{Qualitative Analysis Method}
After completing the user study, participants responded to five open-ended questions addressing comparisons with conventional tools, the effect of animation on interpretation, the impact of different viewing angles, potential use cases, and suggested improvements.

Due to the exploratory nature of this qualitative study, we conducted an inductive analysis approach using reflexive thematic analysis \cite{Braun2022reflexivethematic,braun2024critical} by two coders. The coders initially familiarised themselves with the participants' responses. They then independently coded the responses. In a separate session, they discussed their codes to identify shared ones, and any disagreements were resolved by a third researcher who is a crime scene reconstruction expert among the authors (\#1 in \autoref{tab:experts_details}). They further identified initial themes, then reviewed and refined them to reach a final set of themes.\footnote{ A report on the identified codes, grouping and generated themes is provided in the supplemental materials.} The generated themes, reflexive analysis and a visualisation of quantified codes are detailed in \autoref{sec:qualitative_feedback}.

\subsection{Participants}
\autoref{tab:demographic-participant-info-raw} details the participants' demographic and experience information. We recruited 18 participants, including 6 who had a degree or at least formal training in criminology or crime scene investigation (trained group). The untrained group comprises university students, ranging from undergraduate to PhD level, in fields such as IT, engineering, and science. We denote participant ID (PID) of the trained group with T1 to T6, and untrained group with U1 to U12 in the rest of the paper.

\begin{table}[ht]
\centering
\caption{Participants' demographics and experience. The trained group comprises T1–T6; the untrained group comprises U1–U12. XR experience levels are Daily, Weekly, Monthly, Rarely, and Never. CSI experience levels are Expert, Advanced, Intermediate, Basic, and No knowledge.}
\begin{tabular}{c c c c c}
\hline
PID & Age (Year) & Gender & XR Experience & CSI Experience \\
\hline
T1  & 25 & F & Rarely & Advanced \\
T2  & 52 & M & Monthly & Expert \\
T3  & 47 & M & Rarely & Advanced \\
T4  & [36-45] & M & Rarely & Intermediate \\
T5  & 28 & M & Rarely & Basic \\
T6  & [36-45] & M & Never & Basic \\
\hline
U1  & 23 & M & Rarely & Basic \\
U2  & 28 & M & Monthly & Basic \\
U3 & 35 & F & Rarely & No Knowledge \\
U4 & 34 & F & Rarely & Basic \\
U5 & 39 & M & Never & Intermediate \\
U6 & 33 & F & Never & No Knowledge \\
U7 & 30 & F & Never & No Knowledge \\
U8 & 29 & M & Weekly & Basic \\
U9 & [26-35] & F & Rarely & Basic \\
U10 & [36-45] & F & Weekly & Basic \\
U11 & 28 & F & Monthly & No Knowledge \\
U12 & [26-35] & F & Rarely & Basic \\
\hline
\end{tabular}
\label{tab:demographic-participant-info-raw}
\end{table}

Interestingly, some participants from the untrained group, believed they had basic to intermediate knowledge, primarily because of movies, games, and documentaries they had watched or played.

Before starting stage 1, participants received training on the user interface, \VR navigation, and usage of the tool. For stage 2, participants were instructed to add tracks, create arbitrary props and effects, record and play back animations, as well as sort, trim, and delete slots. They completed these tasks at their own pace under the guidance of a facilitator. This study was approved by Monash University's ethics committee (No. 44402).

\subsection{Setup}
We used the Meta Quest Pro \VR headset for this study. Due to the processing complexity of Gaussian Splatting rendering for the captured building, we ran the application within Unity on a PC and streamed it to the headset via Meta Quest Link (Air Link) over a 5G Wi-Fi network. The application was run on Unity version 2022.3.29f1, using the Mixed Reality Toolkit 3 and the Meta Movement SDK. The PC specifications were: Intel i9-9900K 3.6 GHz CPU, 64 GB RAM, and an Nvidia RTX 2080 Super GPU.

The environment has been scanned from three different locations within the building by Niantic's Scaniverse application on an Apple iPhone 15 Pro Max. The point-cloud data was merged and rendered by Unity Gaussian Splatting package.\footnote{\url{github.com/aras-p/UnityGaussianSplatting}, Access 9-6-2025} The user study was conducted in an indoor space measuring $6 \times  4$ metres, allowing for free movement within the area. We visualised the results data using Matplotlib \cite{Hunter2007} and Seaborn \cite{Waskom2021} libraries.

\section{Quantitative Results}\label{sec:results}
We report on usability, task load, qualitative feedback and observed patterns of behaviour. 

\subsection{Usability and Workload Analysis}\label{sec:usability_workloadanalysis}
The trained expert group was small (n = 6). A Shapiro-Wilk test rejected the normality of Stage 1  SUS  scores~\cite{shapirowilk1965} (\textit{W} = 0.8901, \textit{p} = 0.039). We therefore use Wilcoxon Signed-Rank test~\cite{Wilcoxon1992} and Mann-Whitney U test~\cite{Whitney1947test}---classically non-parametric tests used to analyse questionnaire data in \HCI \cite{Wobbrock2016nonparametric}---where appropriate.\footnote{The SUS and NASA-TLX questionnaires' results as well as comprehensive normality check and justification for tests are provided in the Supplemental Material
.}
We use p-values in this section, but avoid dichotomy in interpretation\footnote{Following current guidelines on statistical interpretation~\cite{amrhein2019scientists,besancon:hal-03342756,Cockburn,Helske}}. We report confidence intervals for overall scores to interpret effect size visually~\cite{stats_chi,dragicevic:hal-01377894}. We discuss here the results for which we have evidence---even if weak---for comparing groups, while full results are available in
the Supplemental Material.

\autoref{fig:sustlx-total-comparison} shows scores between stages and groups for both SUS and NASA-TLX.
Comparing between stages for each group separately using the same test indicates only strong evidence of SUS differences for the untrained group (\textit{W} = 9.5, \textit{p} = 0.042), and weak evidence of differences of NASA-TLX for the trained group (\textit{W} = 0.0, \textit{p} = 0.062) between Stage 1 and 2. The large error bars in the group-separated measures may explain the weaker evidence compared to the combined results presented earlier.

Using Mann-Whitney U test to compare the groups, we found no clear (or only weak) evidence of differences between the trained and untrained groups across any stages or measures. Therefore, we present the detailed scores for combined groups below.

\begin{figure}
    \centering
    \includegraphics[width=\linewidth]{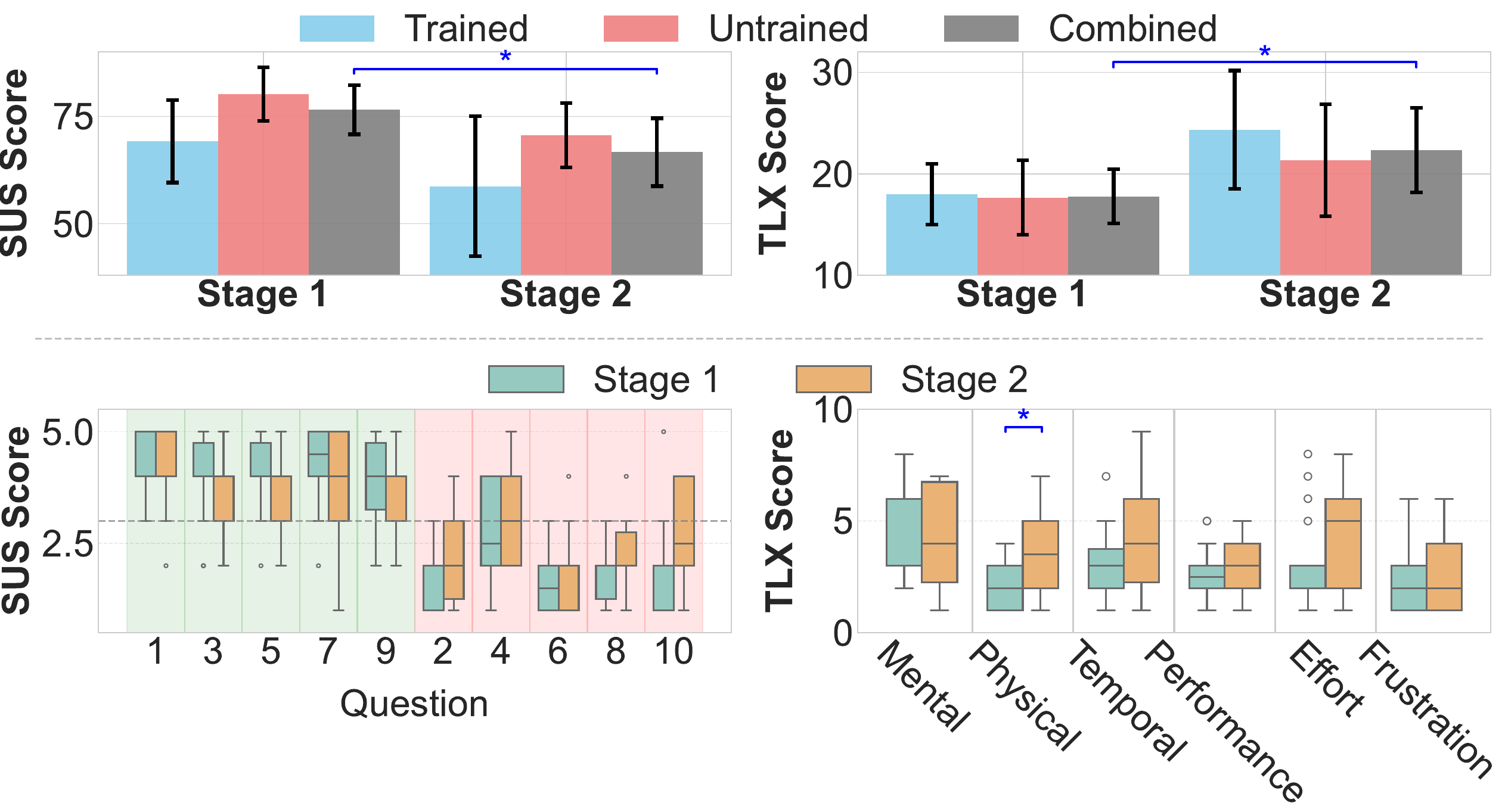}
    \caption{SUS and NASA-TLX scores. Comparison of usability (top left) and task load (top right) for each stage and group. Error bars: 95\% bootstrap CIs. (Bottom left) Comparison of SUS questions' scores between Stage 1 and Stage 2. 
    Higher scores on positive questions and lower scores on negative questions indicate better usability. (Bottom right) NASA-TLX scores between Stage 1 and Stage 2. High scores mean higher demand for all dimensions, except performance where higher score indicates poorer performance. ``*'' indicates significant differences (\textit{p} \textless 0.05) between pairs.}
    \Description{The figure contains four charts displaying usability (SUS) and workload (TLX) scores across two stages of a study. Top-left chart: A grouped bar chart showing total SUS scores for Stage 1 and Stage 2. Each stage includes three bars for trained, untrained, and combined participants, with values ranging from 50 to 80. Error bars indicate variability, with the trained group in Stage 2 showing a particularly wide error bar. Generally, SUS scores are lower in Stage 2. An asterisk above the combined groups' bars indicates a statistically significant difference between stages. Top-right chart: A grouped bar chart showing total TLX scores for Stage 1 and Stage 2, with bars for trained, untrained, and combined participants ranging from 15 to 25. Error bars are shown for each bar, with the trained and untrained groups in Stage 2 having relatively wide error bars. TLX scores generally increase in Stage 2. An asterisk above the combined groups' bars indicates a statistically significant difference. Bottom-left chart: A boxplot of SUS scores for all participants combined, shown for 10 SUS questionnaire items across Stage 1 and Stage 2. Q7 and Q10 show a greater variance across stages. Bottom-right chart: A boxplot of TLX dimensions (Mental, Physical, Temporal, Performance, Effort, Frustration) for all participants combined in Stage 1 and Stage 2. The Physical workload dimension is marked with an asterisk, indicating a statistically significant difference between stages. Temporal and effort dimensions also show greater variance across stages.}
    \label{fig:sustlx-total-comparison}
\end{figure}

The detailed SUS scores (bottom left in \autoref{fig:sustlx-total-comparison}) indicate participants are mostly positive about most aspects of the usability, but need training  (strong evidence of differences for Q7 (\textit{W} = 0.0, \textit{p} = 0.015) and Q10 (\textit{W} = 5.0, \textit{p} = 0.002)) for Stage 2. 

As outlined in \autoref{sec:test_rationale}, raw score comparisons across different domains and cohorts are less informative, but indicate how our prototype has been generally appreciated. Based on the Sauro-Lewis Scale~\cite{lewis2018benchmark}, Stage 1's SUS (\textit{M} = 76.53, \textit{SD} = 12.46) ranks B (above average, 70-79 percentile), and Stage 2's SUS (\textit{M} = 66.67, \textit{SD} = 17.21) as C (average, 41-59 percentile).

\subsection{Stage 1 Descriptive Responses}
\textbf{From witness area.} 
From the end of the hallway most observers (12/18) could identify all purple character movements within their visual field, except for the action of throwing away rubbish, and the hand movements of the green character. They noticed that a fight occurred, but the intention was unclear to them. From that distance, most participants (15/18) failed to identify the weapon held by the green character; among these, some (6/18) even perceived the green character's hand as empty. While all participants regarded the purple character as likely responsible, many (8/18) clearly expressed uncertainty, particularly because of the purple character's posture of surrender prior to the fight. Nevertheless, all participants expressed a degree of uncertainty as they were unable to see inside the room clearly, and were also unsure about the number of individuals present. However, only a few correctly recognised which specific parts the witness could not have seen, because most participants focused primarily on the purple character or did not closely follow exactly when the witness turned around or track their perspective accurately. The trained group's descriptions clearly included more detailed information, such as an estimated distance of the witness from the purple character, \textit{``approximately 10 metres''} (T2), or \textit{``an instrument, probably 60-70 centimetres long''} (T3). The actual distance at their closest point, however, is around 17.5 metres.

\noindent\textbf{Top-down view.} 
A large number of participants (14/18) concluded that the purple character did not perform any direct action against the green character. However, it was still unclear what activity was occurring when the purple character entered the room, which might have triggered the aggressive behaviour of the green character. Nevertheless, the green character's aggressive posture towards the purple character, combined with the purple's surrender posture, led most participants to conclude that the purple character was likely acting in self-defence. Other uncertainties arising from this perspective included the exact weapon held by the green character---although some participants made guesses about it---as well as the precise action the green character was performing. Even though we positioned this angle at a height of 20 metres to provide coverage of the entire corridor, some participants requested to zoom in; upon doing so, they were able to identify the type of weapon held by the green character but failed to recognise the purple character disposing of rubbish. Very few participants (3/18) continued to pay attention to the witness from this angle. However, those participants did not obtain any additional information.

\noindent\textbf{Inside the room.} 
From this angle, all participants had a clear and accurate interpretation of the event and could correctly identify who was the attacker and who was defending. However, some participants were still uncertain about the intention of the green character, questioning why they engaged in the conflict that probably some details have not been visualised. Furthermore, a small number added additional points. For example, comments were made such as, \textit{``I'd be moving a lot faster because that knife is still in play. I wouldn't be dropping the bat''} (T3), \textit{``it doesn't seem premeditated or anything, because why would you leave the bat?''} (T6), and \textit{``the knife angle was a bit out there for actual stab attempt''} (T3). Such observations raised suspicions about whether the self-defence was genuinely provoked by a threat. Another comment was, \textit{``I didn't see any rubbish in the purple character's hand''} (U9) referencing animation details caused uncertainty for the participant about whether the purple character was picking up or placing something that led the green character to initiate the conflict. Nevertheless, this participant considered those possibilities very unlikely.

\subsection{Stage 2 Animation Authoring Performance}\label{sec:animauthor-performance}

\begin{figure}
    \centering
    \includegraphics[width=\columnwidth]{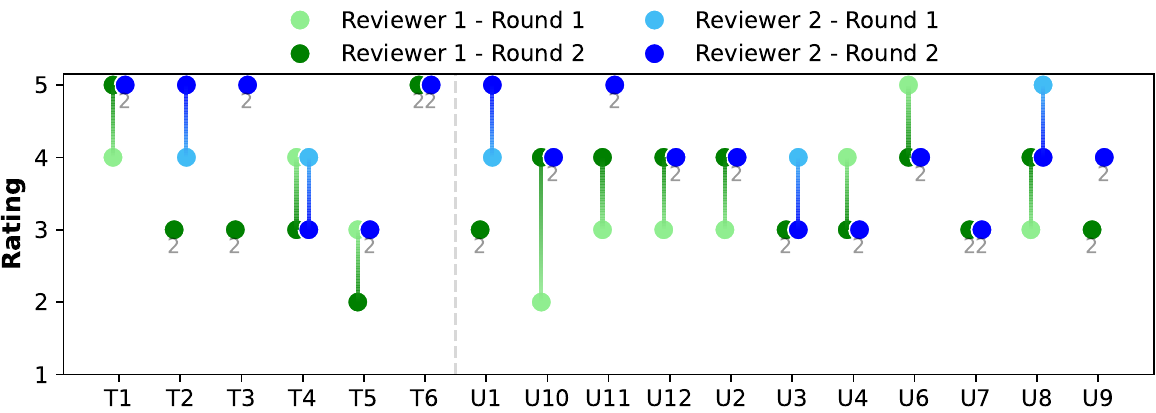}
    \caption{Reviewers' ratings for each participant's animation and changes between rounds. 
    The number next to the points indicates the number of overlapping points.}
    \Description{The figure shows reviewers' ratings for participants' animations across two rounds. The x-axis lists participants as T1–T6 for trained group, separated by a dashed line, then U1–U12 for untrained group. The y-axis shows the rating scale from 1 (lowest) to 5 (highest). Four sets of points are plotted: Reviewer 1 – Round 1 , Reviewer 2 – Round 1, Reviewer 1 – Round 2, and Reviewer 2 – Round 2. Vertical lines connect ratings for each reviewer between rounds to highlight if there is a difference between rating from a reviewer. Ratings vary across participants, with some showing consistent values between rounds and others showing changes of one or two points. Only one participant, U10, received a rating from a single reviewer, with a difference of two points. Several participants have overlapping points where both reviewers gave the same rating in a round, indicated by the number "2" next to the point.}
    \label{fig:rating_participant}
\end{figure}

Reviewers' ratings of participant animations are shown in \autoref{fig:rating_participant}.\footnote{U5's animation was not correctly stored and therefore not rated.} 53\% of  ratings are completely consistent between rounds (18 overlapping points), 44\% (15) differ by one score, and 3\% (1 pair, U10:Reviewer 1) differ by two scores. While some variation in scores is expected, we looked closer at the reasons behind the discrepancies in U10's ratings. Reviewer 1 noted in both rounds that the purple character moves slightly faster and fails to grasp the bat on the first attempt. However, while they initially mentioned in the first round that the purple character \textit{``strikes three times''} instead of two, they did not mention this in the second round. This could be due to a different interpretation or a simple human error. Aside from this case, the rest of the ratings remain largely consistent.

\autoref{fig:rating_participant} suggests the trained group (left side of dashed line) performed slightly better at replicating the animations than the untrained group (right side). 
The primary reason for mistakes in the trained group is related to grasping the bat, with some cases also resulting in timing issues during the fighting part. In contrast, the main issues for the untrained group include grasping the bat, timing, positioning, and the number and the method of strikes performed.

\subsection{Location and Gaze Patterns}
\begin{figure}[ht]
    \centering
\includegraphics[width=\columnwidth]{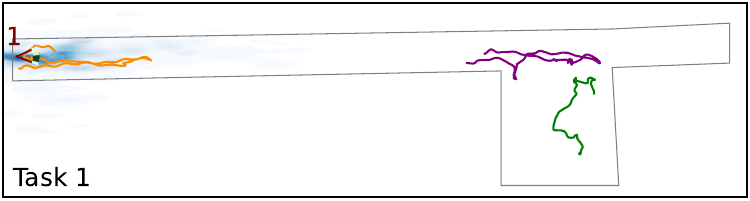}
\includegraphics[width=\columnwidth]{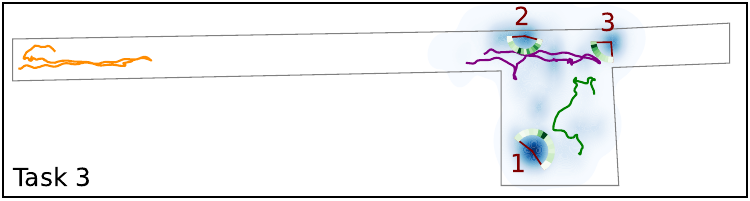}
    \caption{Participants' location distribution during each Task 1 and 3 (with normalised task duration time for all participants). The orange, purple, and green lines represent the positions of the witness, defender, and attacker throughout the animation, respectively. The arcs illustrate the distribution of participants' gaze directions within a 0.5-metre radius from the centre of the numbered clusters (after removing outliers). Cluster 1 in Task 1 shows participants mostly stayed near the door behind the witness to see all characters together, and hence had a narrow range of view directions. They also walked with the witness (orange) to view from their position. Similarly, Clusters 1-3 in Task 3 reveal three main viewing positions. Cluster 3 provides the optimal (narrower) range and visibility of the witness with a smoother distribution of viewing angles ((indicated by the density of green on the arc), while Cluster 1 requires extensive head and eye movement and obscures the witness. Despite this, participants spent more time in Cluster 1, suggesting that finding the optimal position is not always obvious.}
    \Description{The figure consists of two images showing participants' location distributions and gaze directions during Task 1 (top) and Task 3 (bottom). In both tasks, the layout features a narrow horizontal corridor and a study area attached near the lower right. Three paths, represented in different colours, correspond to the movements of the witness, defender, and attacker. The witness path runs near the left end of the corridor, the defender path is located near the junction with the study area, and the attacker path is within the study area. A blue heatmap indicates a density map of participants' positions during the study for both tasks. Small arcs around the densest areas of the density map represent participants' gaze directions at those locations. In Task 1, there is a single high-density area (darker blue), labelled "1", near the door behind the witness, with a narrow range of viewing directions. In Task 3, three high-density areas, labelled 1–3, are present at different corners of the study area: area 2 and 3 are associated with the corridor (2 on the left, 3 on the right), and area 1 is at the bottom left of the study area. The viewing angle arcs vary from narrowest to widest as follows: area 3, area 2, and area 1, with area 3 showing the smoothest distribution of viewing angles.}
    \label{fig:kdeplot-locations}
\end{figure}

\autoref{fig:kdeplot-locations} shows predominant position and gaze directions during tasks 1 to 3. 

\noindent\textbf{Task 1.} A darker area at the far end of the hallway (cluster 1) shows participants' choice for observing the witness's movements, despite being further from the crime scene. The arc overlaid on the positional data also shows a narrow range for gaze direction, \ie participants did not look around much. Some participants moved within this area, closely approaching the witness. 

\begin{figure}[ht]
    \centering
    \includegraphics[width=\columnwidth]{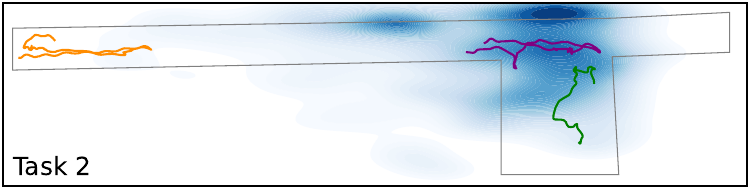}
    \caption{The distribution of gaze hit on the scene from a top-down view during Task 2 from all participants (with normalised task duration time for all participants). The density map shows a strong focus on the victim (purple) path, primarily around the fighting area, with less attention on the attacker (green) and even less on the witness (orange). Since this represents head gaze positions rather than eye gaze, there is a slight offset between the areas of focus and the regions with the highest density of gaze hits.}
    \Description{The figure shows participants' gaze directions mapped onto a floor plan. The layout features a narrow horizontal corridor and a study area attached near the lower right. Three paths, represented in different colours, correspond to the movements of the witness, defender, and attacker. The witness path runs near the left end of the corridor, the defender path is located near the junction with the study area, and the attacker path is within the study area. A blue heatmap overlays the layout, with darker blue indicating a higher density of gaze hits aggregated across all participants. The highest gaze densities appear near the defender path, with slightly lower densities around the attacker path.}
    \label{fig:kdeplot-gazedist}
\end{figure}

\noindent\textbf{Task 2.} Position patterns are not shown as  the entire space is visible from the top-down perspective. We instead analyse the distribution of gaze hit on the scene (see \autoref{fig:kdeplot-gazedist}). It shows that the participants' focus was primarily on the purple and green characters, while significantly less attention was given to the witness.

\noindent\textbf{Task 3} The position distribution in \autoref{fig:kdeplot-locations} indicates that participants moved across the entire room space, but stopped more frequently at three areas (cluster 1 to 3). 
The arcs show that participants positioned near cluster 1 had a wider range of gaze directions with more dispersed gaze angles. They had to repeatedly turn their heads to better track the movements and timing of both characters. In contrast, individuals standing in the hallway (cluster 2 and 3), especially near cluster 3, did not need to make such efforts, as their gaze covered a narrower and more continuous range of angles, which likely resulted in fewer distractions.

\begin{figure}[ht]
    \centering
    
    \includegraphics[width=\columnwidth]{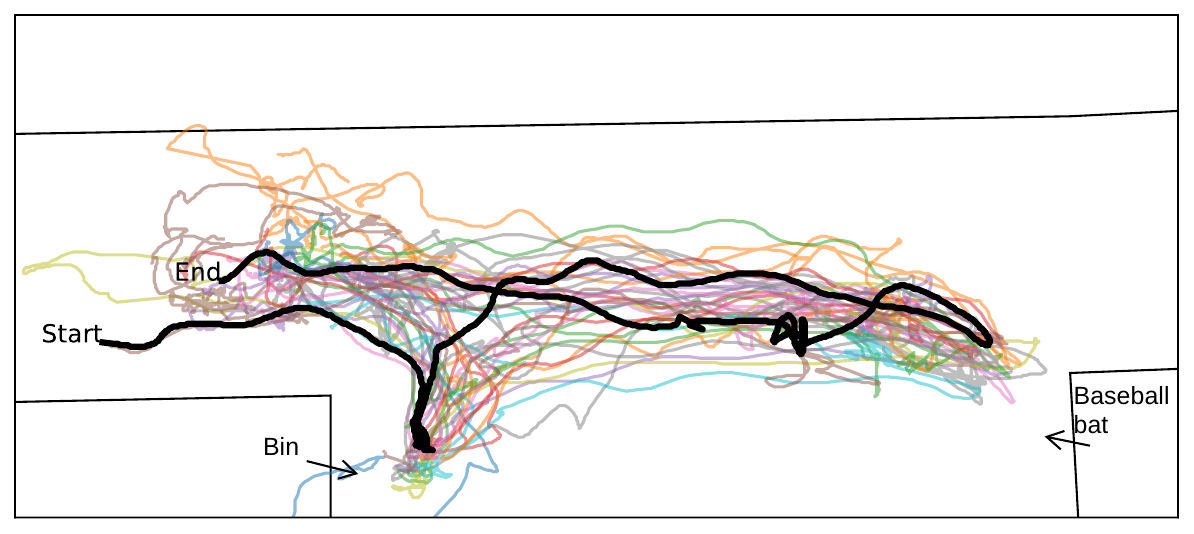}
    \caption{Movement paths of participants in Stage 2 are represented in different colours. The black line shows the movement of the original character animation. This map shows that, while there are similar patterns in the movements of each participant---especially around key areas such as the start and end points, the rubbish bin, and near the baseball bat---accuracy decreases in other areas where there are no clear reference points.}
    \Description{This figure shows the movement map of Stage 2: many thin, coloured paths go from a labelled Start to End across the corridor and the attached study room. A thick black path shows the original animation. Participants' paths cluster near landmarks, such as rubbish bin and baseball bat, and spread out in other areas.}
    \label{fig:participants_movement_stage2}
\end{figure}

\noindent\textbf{Stage 2}. \autoref{fig:participants_movement_stage2} shows movement paths for all participants while creating animation, compared to the path of the purple character from the original animation coloured in black. Considerable variation in the paths taken by participants could potentially influence the event's visibility from the witness's perspective.

\subsection{Observations from User Interaction}
Participants generally played the animation three to four times from the witness's perspective, twice from the top-down view, and twice from within the room. Some participants who positioned themselves appropriately within the room were even prepared to respond after just one playback. While answering questions after tasks 1 to 3,  participants often used gestures to indicate directions, objects, and locations in \VR instead of verbally describing every detail, as if they and the investigator were both present in \VR.

\section{Qualitative Results}\label{sec:qualitative_feedback}
In this section, we report analysis of participants' feedback---including comparisons with conventional methods, animation, viewpoints, use cases, and required improvements---in two levels of detail: a quantified visualisation and a reflexive thematic analysis.

\subsection{Overview}
\autoref{tab:qualitative_feedback_heatmap} presents a quantified summary of the codes derived from participants' questionnaire responses, organised by categories (columns) and individual participants (coloured cells). Quantification has been done based on agreement between two coders---and a third one in cases of disagreement---regarding whether each code reflected a positive, negative, or mixed sentiment; multiple codes with different sentiment also indicated as mixed sentiment. This table shows positive views about the benefits of viewpoints and animation, which were also rated as superior to current practice. Although participants were successful in creating animations, the table shows many negative comments regarding the user interface (UI). The tool was viewed as useful for analysis, examination, and training.  Views were mixed for the tool in court, as discussed in \ref{sec:reflexive_thematic}(2).

% --- THE TABLE ---
\begin{table*}[ht]
\caption{Coded participants' feedback distribution on different categories. Each cell represents one participant's opinion regarding our method, with participant IDs increasing from left to right and top to bottom. \colorbox{BlueGreen!100}{Blue} cells indicate support for that category, \colorbox{red!50}{red} cells indicate opposition, \colorbox{gray!50}{grey} cells indicate mixed opinion or unclear feedback, and empty cells indicate no feedback.}
\centering
\begin{tabular}{| l | p{1.5cm} | c | c | c | c | c | c |}
    \hline
     & Vs. Current Practice & Animation & Viewpoint & UI Usability & Examination & Court Use & Training \\
    \hline

    Trained & 
    \drawHeatmap{0.5}{1, 2, 1, 1, 1, 1} & 
    \drawHeatmap{0.5}{1, 1, 1, 1, 1, 1} & 
    \drawHeatmap{0.5}{1, 1, 1, 1, 1, 1} & 
    \drawHeatmap{0.5}{0, -1, 1, 1, -1, -1} &
    \drawHeatmap{0.5}{1, 1, 1, 1, 0, 1} &
    \drawHeatmap{0.5}{2, 2, 0, 1, 0, 2} &
    \drawHeatmap{0.5}{1, 2, 2, 2, 2, 2} \\
    \hline
    
    Untrained & 
    \drawHeatmap{0.5}{2, 1, 2, 2,2,1,1,1,2,2,1,2} & 
    \drawHeatmap{0.5}{1,1,1,-1,2,1,1,1,1,1,1,1} & 
    \drawHeatmap{0.5}{1,1,1,1,1,1,1,1,1,1,1,1} & 
    \drawHeatmap{0.5}{-1,-1,0,1,1,2,0,-1,1,1,-1,1} &
    \drawHeatmap{0.5}{1,1,1,1,1,1,1,1,1,2,1,1} &
    \drawHeatmap{0.5}{2,2,1,2,2,2,2,2,2,2,2,2} &
    \drawHeatmap{0.5}{2,2,1,2,2,2,2,1,1,1,2,2} \\
    \hline

\end{tabular}
\label{tab:qualitative_feedback_heatmap}
\end{table*}

\subsection{Reflexive Thematic Analysis}\label{sec:reflexive_thematic}

Our analysis of participants' feedback can be presented in six themes:

\noindent \textbf{1. Spatial and temporal representations support visual cognition and perception but can also be falsely persuasive.} 

3D animated reconstructions of crime scenes in \XR offer a natural walk-through experience which is not possible with traditional approaches such as sketches, photographs and video recordings, and even 3D visualisation on a screen. Interpreting tens or hundreds of images, or multiple videos captured from different angles, can be a highly-demanding cognitive task. In contrast, 3D animations provide a single, coherent environment with no sense of disconnection. This was reflected in participant feedback: they strongly favoured the animated reconstructions over photos and videos, noting that animation and different viewpoints enhanced their spatial awareness and ability to form a holistic understanding of the scenario. The \VR representation has also been found helpful in ``\textit{figuring out the motive/intention of people, as the empathy level is way higher with this VR solution}'' (U2). The animation also indicates the timing of actions, which helps identify the causes and assess the appropriateness of interactions. However, a virtual investigation cannot fully replace visiting a scene in person, where assessing various aspects firsthand is essential.

However, non-experts in criminology may sometimes misinterpret the understandability of a representation with the accuracy of the underlying events, which can make the depiction appear more persuasive than warranted. This was evident in comments from some participants who felt they had gained ``\textit{a more accurate and comprehensive understanding of an event}'' (U12) or ``\textit{the interpretation closer to the truth}'' (U11).

\noindent \textbf{2. Legal acceptability of this method needs a clear boundary for usage.} 

Participants expressed mixed views on the provenance of animated reconstructions. Some see the tool as highly useful for courtroom presentations, especially in helping juries understand complex scenarios. Others questioned this use case as ``\textit{rarely an admissible presentation of evidence to a jury}'' (T5) because there is rarely verifiable proof of many depicted details such as body movements, locations, and actions.

The provenance of a reconstruction is highly dependent on the source data from which it is derived. Our current approach, however, does not differentiate between uncertain and certain parts of the reconstruction, which may lead to misleading interpretations in some areas. This limitation raises concerns regarding its admissibility in courtroom contexts. Consequently, a clearly defined scope of appropriate and responsible use is required for court use cases, and further research and refinement will be necessary to meet legal standards for acceptance.

\noindent \textbf{3. Improved accessibility of crime scene recreation, democratises judicial process.}

Participants generally felt capable of recreating scenes and found it relatively easy to recreate animations. This is also supported by the animation assessment results in \autoref{sec:animauthor-performance}. While the trained group had relatively better recreation, the scores were not significantly lower for the untrained group. The animation authoring ease-of-use reflected in participants' feedback as well. For instance, T1 noted that ``\textit{by the end of it, I'm quite confident in being able to access the tool again independently}''. Both the trained and untrained groups were able to produce animations with the tool, although additional training and experience would likely improve their proficiency. The use of embodied movement in \XR made character-animation generation intuitive. Nonetheless, several usability issues limited participants' effectiveness, although such challenges are to be expected in a prototype. Furthermore, while the trained group required more \XR experience to interact with virtual objects, the untrained group would benefit from basic criminology training to consider actions and scenarios more carefully. Overall, the approach shows promise in enabling non-experts to create animations, thereby supporting more democratic ideation and communication tool compared to current practice. 

\noindent \textbf{4. This tool better supports higher-level/abstract analysis.}

Participants viewed the tool as useful for obtaining a holistic overview of a scenario, incorporating multiple angles, actions, and parties, and for clarifying misunderstandings arising from particular perspectives. They also noted that it does not support close inspection of details or high-fidelity reconstruction, as also seen in . As we anticipated, the tool is therefore better suited to broad, high-level exploration rather than precise, fine-grained analysis; such precision would conflict with the convenience and accessibility aims discussed earlier and is not within the tool's intended scope.

Although this limitation may appear similar to concerns about legal acceptability (Theme 2), it is fundamentally distinct. A representation can draw on clearly verifiable sources of data, such as video footage, whereas any reconstruction necessarily involves some degree of data loss. This loss stems not only from current technological limitations but also from design decisions regarding the desired level of fidelity. Consequently, as participants also reflected, the tool is best understood as providing a holistic sense of the scenario and clarifying its dynamics, rather than supporting precise measurements or detailed analytical tasks. However, future work may also address this aspect in the design choices without adding substantially to the burden of complexity or skill requirements.

\noindent \textbf{5. Flexible platform for a wide range of applications, but best suits presentation/knowledge sharing and education.}

Creating animations intuitively and within a short time frame can benefit many applications. As participants envisioned, the tool could support from obvious use cases such as crime scene analysis by integrating and completing obscured scene sections in a video footage, court presentation (with less disturbing pictures), and police training, to less obvious domains including entertainment, emergency planning and training, and museum experiences. For police training, for example, the tool can make diverse scenarios from around the world accessible anywhere and anytime, particularly for high-stress decision-making exercises. This also underscores VR's potential as a platform for authoring animated 3D stories, with the flexibility to be adapted to other contexts. Consequently, it represents a valuable area in which further advances could yield broad and significant impact.

\noindent \textbf{6. This VR tool fits the application but usability can be a barrier to use.}

Participants generally found the tool easy to use, engaging, and even fun and suitable for the crime scene investigation (reflected in the SUS scores as well \autoref{sec:usability_workloadanalysis}). They also identified several opportunities to further strengthen both its analytical capabilities and its usability. Enhancing scene reconstruction with audio and improving the ease and accuracy of acting for animation creation were seen as ways to deepen analytical potential. In particular, embodied animation authoring for highly physical or dangerous actions may be challenging to implement and needs specially designed techniques.

Regarding usability, participants highlighted a range of improvements enabled by newer technologies: more intuitive interfaces and controls, higher resolution and visual fidelity, and the use of hand or voice commands in place of button-based interactions. Emerging headsets such as the Apple Vision Pro and Samsung Galaxy XR already enhanced many of these aspects significantly, suggesting that future studies could readily take advantage of these advancements.
\section{Discussion}\label{sec:discussion}

While the analysis of quantitative and qualitative results has provided some insights, aspects such as ethical and legal considerations, as well as broader implications of this study need to be mentioned to contextualise our findings and better inform future research and practice.

\subsection{Ethical and Legal Considerations}

The practical use of variations of existing methods within a legal context requires technical justification in accordance with ISO 21043 Forensic Sciences \cite{morrison2025forensicscienceISO}. This validation is critical given the potential for severe consequences impacting judicial outcomes. We thus discuss the ethical and legal considerations associated with our proposed method, and recommend clear boundaries of when and how it can best be applied.

\noindent \textbf{Differentiating the evidence-based and hypothetical animation/movements.} 

Evidence-based explanations and representations are the benchmark in forensic work, but reliable evidence is not always available. Consequently, evidence-based, hypothetical, and mixed reconstructions all have important roles in crime scene investigation. Investigation teams typically distinguish these forms easily, reflected in positive views about examination use cases (\autoref{tab:qualitative_feedback_heatmap}), knowledge sharing, and abstract analysis (Themes 4 and 5 in \autoref{sec:reflexive_thematic}). Although concerns exist about admissibility, this is ultimately for the court to decide on a case-by-case basis, often with agreement between prosecution and defence.

Opinions are more mixed regarding the use of such reconstructions in court as presentation tools for juries. Concerns centre on whether they may introduce bias through their realism and falsely persuasive power. This is due to the``seeing-is-believing'' attitude \cite{schofield2009animating,galves1999not}---a tendency to believe realistic representations---while the representation is hypothetical and uncertain; especially that jurors are not trained to distinguish hypotheses from evidence. Nonetheless, trained participants viewed them as moderately useful in court because the animated format can aid comprehension and support judicial decision-making. We intentionally used humanoid robot avatars rather than hyper-realistic human avatars, balancing reduced bias with improved perception, which is critical in decision-making processes \cite{horvath2024reconstruction}. 

We argue that, as with other forms of presentation---that may include both evidence-based and hypothetical representations---uncertainty should either be verbally communicated to jurors or visually highlighted. We echo the same recommendation of prior work \cite{BAILENSON2006} that the more realistic and advanced a representation is, the more accessible it must be. To support accessibility and fairness, we evaluated our approach with lay participants, not only for observation but also animation recreation. Hypothetical representations can thus be challenged by counter-examples introduced by other parties in a judicial process. Our method is particularly useful for testing possible versus impossible scenarios. While a possible scenario is not necessarily actual, it helps narrow the set of impossible ones and curb unrealistic claims. This is valuable in cases such as when eyewitness descriptions need validation, and previous research shows \XR performs well in such contexts \cite{reichherzer_bringing_2021}. It can also be more ethical, offering useful detail without hyper-realism and reducing the need to revisit crime scenes or distressing imagery, thereby limiting further psychological harm to victims and witnesses.

\noindent \textbf{Uses for demonstration of presumed events, not for determining intent or motive.} 

The modelling can be produced at varying levels of detail and precision, even when supporting evidence exists. Our toolkit is designed for ``rapid reconstruction'' and to be ``accessible'' to users without specialised 3D modelling expertise; therefore, detailed and fully precise reconstruction of movements is not intended. This aligns with the visualised location patterns (\autoref{fig:participants_movement_stage2}) and animation authoring performance (\autoref{sec:animauthor-performance}). Instead, our method supports the demonstration of plausible sequences of events through re-enactment, particularly for intervals between events already established through data-supported crime scene reconstructions. It is also important to recognise that crime scenes rarely contain all information required for definitive conclusions---a limitation not unique to our approach. Consequently, determining intent or motive solely from these reconstructions is not always feasible and intended, but useful for verifying hypotheses aligning with prior work \cite{BUCK2013recons3d,reichherzer_bringing_2021}. In addition to evidence-based or hypothetical reconstruction, we thereby specified the intended precision and level-of-details of the proposed tool to support transparent legal assessments of when and how it should be used.

\noindent \textbf{Legal aspects of data acquisition and sharing.}

Another aspect that can both restrict and support our method, as well as similar 3D-captured representations, is the security and privacy concerns associated with data collection and sharing \cite{ranglov2024reconsbenefitconcern}, along with regulations that vary across jurisdictions. For example, 3D capture in private homes and buildings may be prohibited in some areas, even for law enforcement in criminal cases. Furthermore, sharing such data with third parties is often even more restricted. This limits the potential of our tool as an accessible platform for everyone, due to privacy, security of data, and alternation. Even in jurisdictions with more flexible rules, strict regulations around sharing identifiable data and disclaimers for evidence disclosure---particularly when children are involved---remain in place; as they could potentially putting individuals in danger or creating other risks. Having said that, 3D-modelled representations and animations may be preferred in such cases over traditional methods like photography and videography, making them more ethically and legally acceptable.

\subsection{Design Implications and Recommendations}

While our study is set within a legal context, the challenges are not all unique to this domain. The following aspects also represent our insights of this work on important challenges in broader visualisation and HCI research.

\noindent \textbf{Data Provenance. }

As highlighted by participants' feedback in Theme 2 (\autoref{sec:reflexive_thematic}), providing data provenance is critical for improving the transparency and reliability of visualisations in sensitive applications. In our study, the scenario was hypothetical, but our system also supports including photos as thumbnails within the scene. For critical applications, it is essential to clearly indicate the data source and, if the data are synthesised, to explain the generation process. Standard visualisations often incorporate source citations and, in some cases, more detailed information through tooltips. We argue that an interactive 3D environment in \XR can enhance this aspect by enabling users to access detailed data and its sources on demand, and in intuitive manner inspired by tootltips. Traditional methods, such as 3D-rendered videos or other non-interactive representations, lack this capability, and attempting to display all sources can result in visual clutter. In an interactive 3D environment---particularly in \XR---photo thumbnails from the point of capture, as well as videos along with their capture path (\eg by pose estimation of body-worn camera in 3D \cite{flight2022determining}), can effectively convey provenance without overwhelming the user. Nonetheless, effective and possibly automated ways of linking elements in a 3D environment to its underlying data provides opportunities for future research.

\noindent \textbf{Uncertainty Visualisation. }

Based on Themes 2 and 4 (\autoref{sec:reflexive_thematic}), we recognise that legal considerations require clarity about the provenance of information shown in visualisations, i.e. whether there is direct evidence or whether it is inferred. Uncertainty could be informed verbally, or through visualisation. Effective uncertainty visualisation is an active research area in HCI and the visualisation in particular~\cite{jena2020uncertainvissurvey}. Standard charts often use error bars or confidence levels (as in~\autoref{fig:sustlx-total-comparison}), but while such cues support interpretation, they may also raise cognitive load.

Communicating uncertainty is even more challenging in non-standard visualisation media such as 3D or animation. Humanoid robot avatars (as opposed to photorealistic human character avatars) in our application help reduce bias, for example, avoiding identifying features or facial expressions which may induce bias. Other approaches to indicate imprecise or unknown position, movement or appearance in crime scene reconstruction include abstract figures (\eg cones or spheres instead of bodies with limbs) \cite{schofield2009animating}, selectively blurring scene elements, or animated vibration~\cite{Brown2004visualvibration}. Despite the benefits of these techniques, they may also overload and confuse audiences~\cite{fiedler2003eyedeceiving}, especially non-experts like juries. The relative effectiveness of these techniques and their combination, however, warrants further studies.

\noindent \textbf{Expert-Centred Co-Design. }

Our ongoing collaboration with forensic investigation experts offers lessons for future researchers aiming to conduct co-design requiring multiple, sustained sessions with domain specialists. In brief:

\begin{enumerate}
    \item In expert co-design, the focus should be on the depth of insights rather than participant numbers. Smaller sample sizes that yield richer insights are increasingly accepted in the HCI community; for instance, the recent workshop ``Engaging Critical Workforce In co-Design and Assessment'' at IEEE VIS 2025 \footnote{https://sites.google.com/view/ecwidna, Accessed 18-11-2025} discussed this matter in depth, and it has been done in other HCI publications \cite{Caine2016samplesize,besancon:hal-03012861,pooryousef2023working}.
    \item In sensitive domains, ethics approvals from one institution may not be recognised elsewhere, and non-research-focused institutions may process applications slowly. Flexible study designs in ethics applications are therefore recommended, not postponing to final version of study design, though this does not apply to clinical trials or high-risk studies due to potential misuse \cite{frank2023ethics}.
    \item Involving highly experienced experts as co-authors---rather than only as participants---is highly valuable. This encourages deeper engagement, enriches design outcomes, and can address ethical applications issues. When transparently reported, as in prior studies \cite{besancon:hal-03012861,pooryousef2023working}, this practice is generally acceptable unless it introduces significant risk of bias. We followed this approach in this paper.
    \item Using higher-fidelity prototypes, such as via a Wizard-of-Oz approach, instead of common paper-based co-design sessions or workshops enhances communication between domain experts and HCI researchers, particularly for emerging technologies like \XR. Without such prototypes, discussions often remain at a basic level due to experts' unfamiliarity with the emerging technologies. While this requires more researcher effort, it saves experts' time and improves overall outcomes.
\end{enumerate}

\subsection{Limitations}

Experts were difficult to recruit due to security restrictions at these sensitive institutions, leaving only a small pool of experts for our evaluation, limiting statistically robust analysis. However, methodologists and HCI researchers have argued that there is no cut-off number that would dramatically improve the credibility of HCI-like quantitative analyses~\cite{Bacchetti2010samplesize}, and most HCI and visualisation studies rely on a few participants ~\cite{Caine2016samplesize}. This is all the more true when it comes to experts recruiting that is both difficult, expensive, and an inherent limitation of our research work with experts that we are used to deal with and interpret accordingly \cite{besancon:hal-03012861,Pooryousef2025Thesis}.  We also acknowledge that our current work involves only two main stakeholder groups, and does not include other key parties such as judges, lawyers, and policymakers. Although expert participants could anticipate some perspectives from these groups based on their familiarity with the field, they cannot fully assess considerations that only domain specialists would recognise.

We also faced several technological challenges. The Meta Movement SDK does not provide precise lower-body joint positioning, limiting accuracy of character animations for lower-body movements. In the long run, markerless tracking and video-to-animation methods using advanced computer vision algorithms \cite{Ihara2025Video2MR,zheng2023bodytrack,Liu2022posesurvey} will enhance full-body motion capture and potentially enable direct conversion of CCTV and body-worn camera footage \cite{flight2022determining} into character animations. However, there will always be need for manual animation in hypothesis testing when footage is unavailable. In addition, we did not have access to professional scanners---which may also not be an affordable option in low-resource countries---and this limitation may have affected participants' perceptions.
\section{Conclusion and Future Work}\label{sec:conclustion}

Through a co-design process with forensic experts, we developed a framework and toolkit with a prototype implementation called ``Criminator''. Criminator enables users to reconstruct and re-enact animated crime scenes in \VR, without needing conventional modelling or animation skills. We evaluated Criminator with both crime scene investigators and laypeople, with results indicating relative success in scene observation and animation authoring tasks. Although animation creation was found to be slightly more challenging than observation, there were few significant differences in performance, evidencing the tool's ease of use. Expert participants also highlighted potential uses for hypothesis testing---such as event sequencing and witness viewpoints---courtroom presentations, and police training. Given rapid advancements in computer vision and 3D scanning, such tools could soon become even more practical. 

In future research, it is essential to expand the participant groups and involve other key stakeholders such as judges and lawyers in the design and evaluation process, to better identify limitation, concerns, and additional benefits raised by such tools. There is also a need for more robust comparative and empirical studies, investigating newer devices capable of enabling a seamless hybrid desktop-XR approach, as well as incorporating video-to-animation algorithms.

\begin{acks}
The authors wish to thank Janis Dalins and Campbell Wilson (both from Monash University) for their contribution to the co-design process, and all of the participants of the conducted study for their insightful feedback and comments. This work was funded by the Victorian Institute of Forensic Medicine (VIFM) and partially supported by the Knut and Alice Wallenberg Foundation (KAW 2019.0024) and the Marcus and Amalia Wallenberg Foundation (MAW 2023.0130).
\end{acks}

\bibliographystyle{ACM-Reference-Format}
\bibliography{Refs}

@techreport{OSAC2025-N-0004,
	title        = {Standard Criteria for Crime Scene Reconstruction},
	author       = {Crime Scene Investigation \& Reconstruction Subcommittee},
	year         = 2025,
	month        = {March},
	number       = {OSAC 2025-N-0004},
	url          = {https://www.nist.gov/system/files/documents/2025/03/03/OSAC%202025-N-0004-Standard%20Criteria%20for%20Crime%20Scene%20Reconstruction_OPEN%20COMMENT%20VERSION%201.0.pdf},
	institution  = {Organization of Scientific Area Committees (OSAC) for Forensic Science},
	type         = {Draft OSAC Proposed Standard},
	version      = {1.0}
}

@article{besancon:hal-03012861,
  TITLE = {{The State of the Art of Spatial Interfaces for 3D Visualization}},
  AUTHOR = {Besan{\c c}on, Lonni and Ynnerman, Anders and Keefe, Daniel F and Yu, Lingyun and Isenberg, Tobias},
  URL = {https://inria.hal.science/hal-03012861},
  JOURNAL = {{Computer Graphics Forum}},
  PUBLISHER = {{Wiley}},
  VOLUME = {40},
  NUMBER = {1},
  PAGES = {293--326},
  YEAR = {2021},
  MONTH = Feb,
  DOI = {10.1111/cgf.14189},
  HAL_ID = {hal-03012861},
  HAL_VERSION = {v1},
}

@inproceedings{besancon:hal-01436206,
	title        = {Mouse, Tactile, and Tangible Input for 3D Manipulation},
	author       = {Besan{\c c}on, Lonni and Issartel, Paul and Ammi, Mehdi and Isenberg, Tobias},
	year         = 2017,
	month        = {May},
	booktitle    = {Proceedings of the ACM Conference on Human Factors in Computing Systems (CHI)},
	address      = {Denver, United States},
	pages        = {4727--4740},
	doi          = {10.1145/3025453.3025863},
	url          = {https://inria.hal.science/hal-01436206},
	hal_id       = {hal-01436206},
	hal_version  = {v1}
}

@article{sereno:hal-03699232,
	title        = {Hybrid Touch/Tangible Spatial Selection in Augmented Reality},
	author       = {Sereno, Mickael and Gosset, St{\'e}phane and Besan{\c c}on, Lonni and Isenberg, Tobias},
	year         = 2022,
	month        = {Jun},
	journal      = {Computer Graphics Forum},
	publisher    = {Wiley},
	volume       = 41,
	number       = 3,
	pages        = {403--415},
	doi          = {10.1111/cgf.14550},
	url          = {https://inria.hal.science/hal-03699232},
	hal_id       = {hal-03699232},
	hal_version  = {v1}
}

@article{amrhein2019scientists,
	title        = {Scientists rise up against statistical significance},
	author       = {Amrhein, Valentin and Greenland, Sander and McShane, Blake},
	year         = 2019,
	journal      = {Nature},
	publisher    = {Nature Publishing Group UK London},
	volume       = 567,
	number       = 7748,
	pages        = {305--307}
}

@inproceedings{stats_chi,
	title        = {The Continued Prevalence of Dichotomous Inferences at CHI},
	author       = {Besan\c{c}on, Lonni and Dragicevic, Pierre},
	year         = 2019,
	booktitle    = {Extended Abstracts of the 2019 CHI Conference on Human Factors in Computing Systems},
	location     = {Glasgow, Scotland Uk},
	publisher    = {Association for Computing Machinery},
	address      = {New York, NY, USA},
	series       = {CHI EA '19},
	pages        = {1--11},
	doi          = {10.1145/3290607.3310432},
	isbn         = 9781450359719,
	url          = {https://doi.org/10.1145/3290607.3310432},
	numpages     = 11
}

@incollection{dragicevic:hal-01377894,
	title        = {Fair Statistical Communication in HCI},
	author       = {Dragicevic, Pierre},
	year         = 2016,
	booktitle    = {Modern Statistical Methods for HCI},
	publisher    = {Springer},
	pages        = {291--330},
	doi          = {10.1007/978-3-319-26633-6\_13},
	url          = {https://inria.hal.science/hal-01377894},
	hal_id       = {hal-01377894},
	hal_version  = {v1}
}

@inproceedings{besancon:hal-01795744,
	title        = {Reducing Affective Responses to Surgical Images through Color Manipulation and Stylization},
	author       = {Besan{\c c}on, Lonni and Semmo, Amir and Biau, David J. and Frachet, Bruno and Pineau, Virginie and Sariali, El Hadi and Taouachi, Rabah and Isenberg, Tobias and Dragicevic, Pierre},
	year         = 2018,
	month        = {Aug},
	booktitle    = {Proceedings of the Joint Symposium on Computational Aesthetics, Sketch-Based Interfaces and Modeling, and Non-Photorealistic Animation and Rendering},
	publisher    = {ACM Press},
	address      = {Victoria, Canada},
	pages        = {4:1--4:13},
	doi          = {10.1145/3229147.3229158},
	url          = {https://inria.hal.science/hal-01795744},
	hal_id       = {hal-01795744},
	hal_version  = {v4}
}

@article{Besancon:2020:RAR,
	title        = {Reducing Affective Responses to Surgical Images and Videos Through Stylization},
	author       = {Besan\c{c}on, Lonni and Semmo, Amir and Biau, David and Frachet, Bruno and Pineau, Virginie and Sariali, El Hadi and Soubeyrand, Marc and Taouachi, Rabah and Isenberg, Tobias and Dragicevic, Pierre},
	year         = 2020,
	journal      = {Computer Graphics Forum},
	volume       = 39,
	number       = 1,
	pages        = {462--483},
	doi          = {https://doi.org/10.1111/cgf.13886},
	url          = {https://onlinelibrary.wiley.com/doi/abs/10.1111/cgf.13886},
	eprint       = {https://onlinelibrary.wiley.com/doi/pdf/10.1111/cgf.13886}
}

@article{Helske,
	title        = {Can Visualization Alleviate Dichotomous Thinking? Effects of Visual Representations on the Cliff Effect},
	author       = {Helske, Jouni and Helske, Satu and Cooper, Matthew and Ynnerman, Anders and Besan\c{c}on, Lonni},
	year         = 2021,
	journal      = {IEEE Transactions on Visualization and Computer Graphics},
	volume       = 27,
	number       = 8,
	pages        = {3397--3409},
	doi          = {10.1109/TVCG.2021.3073466}
}

@unpublished{besancon:hal-03342756,
	title        = {Definitely Maybe: Hedges And Boosters in the HCI Literature},
	author       = {Besan{\c c}on, Lonni and Jansen, Yvonne and Cockburn, Andy and Dragicevic, Pierre},
	year         = 2021,
	month        = {Nov},
	doi          = {10.31219/osf.io/mjg7h},
	url          = {https://inria.hal.science/hal-03342756},
	note         = {working paper or preprint},
	hal_id       = {hal-03342756},
	hal_version  = {v1}
}

@article{Cockburn,
	title        = {Threats of a replication crisis in empirical computer science},
	author       = {Cockburn, Andy and Dragicevic, Pierre and Besan\c{c}on, Lonni and Gutwin, Carl},
	year         = 2020,
	month        = jul,
	journal      = {Commun. ACM},
	publisher    = {Association for Computing Machinery},
	address      = {New York, NY, USA},
	volume       = 63,
	number       = 8,
	pages        = {70--79},
	doi          = {10.1145/3360311},
	issn         = {0001-0782},
	url          = {https://doi.org/10.1145/3360311},
	issue_date   = {August 2020},
	numpages     = 10
}

@inproceedings{poelman_as_2012,
	title        = {As if being there: mediated reality for crime scene investigation},
	shorttitle   = {As if being there},
	author       = {Poelman, Ronald and Akman, Oytun and Lukosch, Stephan and Jonker, Pieter},
	year         = 2012,
	booktitle    = {Proceedings of the {ACM} 2012 conference on {Computer} {Supported} {Cooperative} {Work} - {CSCW} '12},
	publisher    = {ACM Press},
	address      = {Seattle, Washington, USA},
	pages        = 1267,
	doi          = {10.1145/2145204.2145394},
	isbn         = {978-1-4503-1086-4},
	url          = {http://dl.acm.org/citation.cfm?doid=2145204.2145394},
	urldate      = {2020-12-13}
}

@inproceedings{pooryousef2023working,
	title        = {Working with forensic practitioners to understand the opportunities and challenges for mixed--reality digital autopsy},
	author       = {Pooryousef, Vahid and Cordeil, Maxime and Besan{\c{c}}on, Lonni and Hurter, Christophe and Dwyer, Tim and Bassed, Richard},
	year         = 2023,
	booktitle    = {Proceedings of the 2023 CHI Conference on Human Factors in Computing Systems},
	pages        = {1--15}
}

@ARTICLE{pooryousef2024AutopsyDoc,
  author={Pooryousef, Vahid and Cordeil, Maxime and Besançon, Lonni and Bassed, Richard and Dwyer, Tim},
  journal={IEEE Transactions on Visualization and Computer Graphics}, 
  title={Collaborative Forensic Autopsy Documentation and Supervised Report Generation Using a Hybrid Mixed-Reality Environment and Generative AI}, 
  year={2024},
  volume={30},
  number={11},
  pages={7452-7462},
  doi={10.1109/TVCG.2024.3456212}}

@article{affolter_applying_2019,
	title = {Applying augmented reality during a forensic autopsy—{Microsoft} {HoloLens} as a {DICOM} viewer},
	volume = {16},
	issn = {22124780},
	url = {https://linkinghub.elsevier.com/retrieve/pii/S2212478018300650},
	doi = {10.1016/j.jofri.2018.11.003},
	language = {en},
	urldate = {2020-12-01},
	journal = {Journal of Forensic Radiology and Imaging},
	author = {Affolter, Raffael and Eggert, Sebastian and Sieberth, Till and Thali, Michael and Ebert, Lars Christian},
	month = mar,
	year = {2019},
	pages = {5--8},
}

@article{frank2023ethics,
  author    = {Frank, Florian and Florens, Nicolas and Meyerowitz-Katz, Gideon and et al.},
  title     = {Raising concerns on questionable ethics approvals – a case study of 456 trials from the Institut Hospitalo-Universitaire Méditerranée Infection},
  journal   = {Research Integrity and Peer Review},
  volume    = {8},
  number    = {1},
  pages     = {9},
  year      = {2023},
  doi       = {10.1186/s41073-023-00134-4},
  url       = {https://doi.org/10.1186/s41073-023-00134-4}
}

@inproceedings{streefkerk_art_2013,
	title        = {The art of csi: {An} augmented reality tool (art) to annotate crime scenes in forensic investigation},
	author       = {Streefkerk, Jan Willem and Houben, Mark and van Amerongen, Pjotr and ter Haar, Frank and Dijk, Judith},
	year         = 2013,
	booktitle    = {Virtual, {Augmented} and {Mixed} {Reality}. {Systems} and {Applications}: 5th {International} {Conference}, {VAMR} 2013, {Held} as {Part} of {HCI} {International} 2013, {Las} {Vegas}, {NV}, {USA}, {July} 21-26, 2013, {Proceedings}, {Part} {II} 5},
	publisher    = {Springer},
	pages        = {330--339},
	doi          = {10.1007/978-3-642-39420-1_35},
	isbn         = {3-642-39419-1}
}

@inproceedings{datcu_handheld_2016,
	title        = {Handheld augmented reality for distributed collaborative crime scene investigation},
	author       = {Datcu, Drago\c{s} and Lukosch, Stephan G. and Lukosch, Heide K.},
	year         = 2016,
	booktitle    = {Proceedings of the 2016 {ACM} {International} {Conference} on {Supporting} {Group} {Work}},
	pages        = {267--276}
}

@inproceedings{gee_augmented_2010,
	title        = {Augmented crime scenes: virtual annotation of physical environments for forensic investigation},
	author       = {Gee, Andrew P. and Escamilla-Ambrosio, P. J. and Webb, Matthew and Mayol-Cuevas, Walterio and Calway, Andrew},
	year         = 2010,
	booktitle    = {Proceedings of the 2nd {ACM} {Workshop} on {Multimedia} in {Forensics}, {Security} and {Intelligence}},
	pages        = {105--110}
}

@inproceedings{tolstolutsky_experience_2021,
	title        = {The {Experience} of {Using} {Augmented} {Reality} in the {Reconstruction} of the {Crime} {Scene} {Committed} in {Transport}},
	author       = {Tolstolutsky, Vladimir and Kuzenkova, Galina and Malichenko, Victor},
	year         = 2021,
	booktitle    = {International {Scientific} {Siberian} {Transport} {Forum}},
	publisher    = {Springer},
	pages        = {1095--1102}
}

@article{albeedan_evaluating_2023,
	title        = {Evaluating the {Use} of {Mixed} {Reality} in {CSI} {Training} through the {Integration} of the {Task}-{Technology} {Fit} and {Technology} {Acceptance} {Model}},
	author       = {Albeedan, Meshal and Kolivand, Hoshang and Hammady, Ramy},
	year         = 2023,
	journal      = {IEEE Access},
	note         = {ISBN: 2169-3536 Publisher: IEEE}
}

@inproceedings{rinaldi_crime_2022,
	title        = {Crime {Scene} {VieweR} ({CSVR}): performing crime scene investigation in {Virtual} {Reality}},
	author       = {Rinaldi, Vincenzo and Hackman, Lucina and Daeid, Niamh Nic},
	year         = 2022,
	booktitle    = {1st {International} {Conference} on {eXtended} {Reality}}
}

@inproceedings{reichherzer_bringing_2021,
	title        = {Bringing the jury to the scene of the crime: {Memory} and decision-making in a simulated crime scene},
	author       = {Reichherzer, Carolin and Cunningham, Andrew and Coleman, Tracey and Cao, Ruochen and McManus, Kurt and Sheppard, Dion and Kohler, Mark and Billinghurst, Mark and Thomas, Bruce H.},
	year         = 2021,
	booktitle    = {Proceedings of the 2021 {CHI} {Conference} on {Human} {Factors} in {Computing} {Systems}},
	pages        = {1--12}
}

@inproceedings{suncksen_preparing_2019,
	title        = {Preparing and {Guiding} {Forensic} {Crime} {Scene} {Inspections} in {Virtual} {Reality}},
	author       = {S\"{u}ncksen, Matthias and Teistler, Michael and Hamester, Frederik and Ebert, Lars C.},
	year         = 2019,
	month        = sep,
	booktitle    = {Proceedings of {Mensch} und {Computer} 2019},
	publisher    = {ACM},
	address      = {Hamburg Germany},
	pages        = {755--758},
	doi          = {10.1145/3340764.3344903},
	isbn         = {978-1-4503-7198-8},
	url          = {https://dl.acm.org/doi/10.1145/3340764.3344903},
	urldate      = {2024-01-10},
	language     = {en}
}

@article{albeedan_designing_2024,
	title        = {Designing and evaluation of a mixed reality system for crime scene investigation training: a hybrid approach},
	shorttitle   = {Designing and evaluation of a mixed reality system for crime scene investigation training},
	author       = {Albeedan, Meshal and Kolivanda, Hoshang and Hammady, Ramy},
	year         = 2024,
	month        = jun,
	journal      = {Virtual Reality},
	volume       = 28,
	number       = 3,
	pages        = 127,
	doi          = {10.1007/s10055-024-01018-8},
	issn         = {1434-9957},
	url          = {https://link.springer.com/10.1007/s10055-024-01018-8},
	urldate      = {2024-06-28},
	language     = {en}
}

@online{flipsidexrMetaverseSocial,
	title        = {{T}he metaverse social media platform built on imagination --- flipsidexr.com},
	note         = {[Accessed 22-07-2024]},
	howpublished = {\url{flipsidexr.com}}
}

@article{kerbl2023gaussian,
	title        = {3D Gaussian Splatting for Real-Time Radiance Field Rendering},
	author       = {Kerbl, Bernhard and Kopanas, Georgios and Leimkuehler, Thomas and Drettakis, George},
	year         = 2023,
	month        = jul,
	journal      = {ACM Trans. Graph.},
	publisher    = {Association for Computing Machinery},
	address      = {New York, NY, USA},
	volume       = 42,
	number       = 4,
	doi          = {10.1145/3592433},
	issn         = {0730-0301},
	url          = {https://doi.org/10.1145/3592433},
	issue_date   = {August 2023},
	articleno    = 139,
	numpages     = 14
}

@article{Sanders2008Cocreation,
	title        = {Co-creation and the new landscapes of design},
	author       = {Elizabeth B.-N. Sanders and Pieter Jan Stappers},
	year         = 2008,
	journal      = {CoDesign},
	publisher    = {Taylor \& Francis},
	volume       = 4,
	number       = 1,
	pages        = {5--18},
	doi          = {10.1080/15710880701875068},
	url          = {https://doi.org/10.1080/15710880701875068}
}

@article{shapirowilk1965,
	title        = {An analysis of variance test for normality (complete samples)},
	author       = {SHAPIRO, S. S. and WILK, M. B.},
	year         = 1965,
	month        = 12,
	journal      = {Biometrika},
	volume       = 52,
	number       = {3-4},
	pages        = {591--611},
	doi          = {10.1093/biomet/52.3-4.591},
	issn         = {0006-3444}
}

@article{lewis2018benchmark,
	title        = {Item benchmarks for the system usability scale},
	author       = {Lewis, James R. and Sauro, Jeff},
	year         = 2018,
	month        = may,
	journal      = {J. Usability Studies},
	publisher    = {Usability Professionals' Association},
	address      = {Bloomingdale, IL},
	volume       = 13,
	number       = 3,
	pages        = {158--167},
	issue_date   = {May 2018},
	numpages     = 10
}

@article{flight2022determining,
	title        = {Determining the accuracy of the 'place in 3D' function in FARO Scene},
	author       = {Flight, Chris and Ballantyne, Kaye N},
	year         = 2022,
	journal      = {J Assoc Crime Scene Reconstr},
	volume       = 26
}

@misc{vizitech_vr_forensics,
	title        = {VR Forensics},
	author       = {Vizitech USA},
	note         = {Accessed: March 17, 2025},
	howpublished = {\url{https://www.vizitechusa.com/walking-with-the-diciples}}
}

@inbook{Woodward2021,
	title        = {Cost-Effective and Eco-Friendly Fire Investigation Training Using Photorealistic Interactive Room Scale Virtual Reality},
	author       = {Woodward, Bradley Friend},
	year         = 2021,
	booktitle    = {XR Case Studies: Using Augmented Reality and Virtual Reality Technology in Business},
	publisher    = {Springer International Publishing},
	address      = {Cham},
	pages        = {147--156},
	doi          = {10.1007/978-3-030-72781-9_18},
	isbn         = {978-3-030-72781-9},
	url          = {https://doi.org/10.1007/978-3-030-72781-9_18},
	editor       = {Jung, Timothy and Dalton, Jeremy}
}

@misc{victoryxr_csi_forensics,
	title        = {CSI Forensics},
	author       = {VictoryXR},
	year         = 2023,
	urldate      = {2025-03-17},
	note         = {Accessed: March 17, 2025},
	howpublished = {\url{https://www.victoryxr.com/csi-forensics/}}
}

@article{Chang28062024,
	title        = {Using Virtual Reality to Enhance Forensic Science Education: Effects on CSI Learning Achievements, Situational Interest and Cognitive Load},
	author       = {Rong-Chi Chang},
	year         = 2024,
	journal      = {Journal of Criminal Justice Education},
	publisher    = {Routledge},
	volume       = {0},
	number       = {0},
	pages        = {1--20},
	doi          = {10.1080/10511253.2024.2372860}
}

@article{Cannavo2019vrcharanim,
	title        = {Immersive Virtual Reality-Based Interfaces for Character Animation},
	author       = {Cannav\`{o}, Alberto and Demartini, Claudio and Morra, Lia and Lamberti, Fabrizio},
	year         = 2019,
	journal      = {IEEE Access},
	volume       = 7,
	number       = {},
	pages        = {125463--125480},
	doi          = {10.1109/ACCESS.2019.2939427}
}

@inproceedings{Vogel2018animvr,
	title        = {AnimationVR - Interactive Controller-based Animating in Virtual Reality},
	author       = {Vogel, Daniel and Lubos, Paul and Steinicke, Frank},
	year         = 2018,
	booktitle    = {2018 IEEE 1st Workshop on Animation in Virtual and Augmented Environments (ANIVAE)},
	volume       = {},
	number       = {},
    publisher   = {IEEE},
    location    = {Reutlingen, Germany},
	pages        = {1--6},
	doi          = {10.1109/ANIVAE.2018.8587268}
}

@inproceedings{Zhou2024timetunnel,
	title        = {TimeTunnel: Integrating Spatial and Temporal Motion Editing for Character Animation in Virtual Reality},
	author       = {Zhou, Qian and Ledo, David and Fitzmaurice, George and Anderson, Fraser},
	year         = 2024,
	booktitle    = {Proceedings of the 2024 CHI Conference on Human Factors in Computing Systems},
	location     = {Honolulu, HI, USA},
	publisher    = {Association for Computing Machinery},
	address      = {New York, NY, USA},
	series       = {CHI '24},
	doi          = {10.1145/3613904.3641927},
	isbn         = 9798400703300,
	url          = {https://doi.org/10.1145/3613904.3641927},
	articleno    = 101,
	numpages     = 17
}

@inproceedings{li2024anicraft,
	title        = {AniCraft: Crafting Everyday Objects as Physical Proxies for Prototyping 3D Character Animation in Mixed Reality},
	author       = {Li, Boyu and Yuan, Linping and Yan, Zhe and Liu, Qianxi and Shen, Yulin and Wang, Zeyu},
	year         = 2024,
	location     = {Pittsburgh, PA, USA},
	publisher    = {Association for Computing Machinery},
	address      = {New York, NY, USA},
	series       = {UIST '24},
	doi          = {10.1145/3654777.3676325},
	isbn         = 9798400706288,
	url          = {https://doi.org/10.1145/3654777.3676325},
	articleno    = 99,
	numpages     = 14
}

@inproceedings{garcia2019spatial,
	title        = {Spatial Motion Doodles: Sketching Animation in VR Using Hand Gestures and Laban Motion Analysis},
	author       = {Garcia, Maxime and Ronfard, Remi and Cani, Marie-Paule},
	year         = 2019,
	booktitle    = {Proceedings of the 12th ACM SIGGRAPH Conference on Motion, Interaction and Games},
	location     = {Newcastle upon Tyne, United Kingdom},
	publisher    = {Association for Computing Machinery},
	address      = {New York, NY, USA},
	series       = {MIG '19},
	doi          = {10.1145/3359566.3360061},
	isbn         = 9781450369947,
	url          = {https://doi.org/10.1145/3359566.3360061},
	articleno    = 10,
	numpages     = 10
}

@inproceedings{chen2023doubledoodles,
	title        = {Double Doodles: Sketching Animation in Immersive Environment With 3+6 DOFs Motion Gestures},
	author       = {Chen, Ruizhao and Pan, Ye and Deng, Zhigang and Wang, Lili and Ma, Lizhuang},
	year         = 2023,
	booktitle    = {Proceedings of the 31st ACM International Conference on Multimedia},
	location     = {Ottawa ON, Canada},
	publisher    = {Association for Computing Machinery},
	address      = {New York, NY, USA},
	series       = {MM '23},
	pages        = {6998--7006},
	doi          = {10.1145/3581783.3613783},
	isbn         = 9798400701085,
	url          = {https://doi.org/10.1145/3581783.3613783},
	numpages     = 9
}

@inproceedings{Quilez2016Quil,
	title        = {Quill: VR drawing in the production of Oculus Story Studio's new movie},
	author       = {Quilez, Inigo},
	year         = 2016,
	booktitle    = {ACM SIGGRAPH 2016 Real-Time Live!},
	location     = {Anaheim, California},
	publisher    = {Association for Computing Machinery},
	address      = {New York, NY, USA},
	series       = {SIGGRAPH '16},
	doi          = {10.1145/2933540.2933549},
	isbn         = 9781450343787,
	url          = {https://doi.org/10.1145/2933540.2933549},
	articleno    = 41,
	numpages     = 1
}

@inproceedings{Dario2018vranim,
	title        = {Anim VR},
	author       = {Seyb, Dario and Grajetzki, Milan and Chin, Grace and Wilkinson, Sasha},
	year         = 2018,
	booktitle    = {ACM SIGGRAPH 2018 Virtual, Augmented, and Mixed Reality},
	location     = {Vancouver, British Columbia, Canada},
	publisher    = {Association for Computing Machinery},
	address      = {New York, NY, USA},
	series       = {SIGGRAPH '18},
	doi          = {10.1145/3226552.3226553},
	isbn         = 9781450358217,
	articleno    = 5,
	numpages     = 1
}

@article{green1985wizard,
	title        = {The Rapid Development of User Interfaces: Experience with the Wizard of OZ Method},
	author       = {Paul Green and Lisa Wei-Haas},
	year         = 1985,
	journal      = {Proceedings of the Human Factors Society Annual Meeting},
	volume       = 29,
	number       = 5,
	pages        = {470--474},
	doi          = {10.1177/154193128502900515},
	url          = {https://doi.org/10.1177/154193128502900515},
	eprint       = {https://doi.org/10.1177/154193128502900515}
}

@incollection{HART1988nasatlx,
	title        = {Development of NASA-TLX (Task Load Index): Results of Empirical and Theoretical Research},
	author       = {Sandra G. Hart and Lowell E. Staveland},
	year         = 1988,
	booktitle    = {Human Mental Workload},
	publisher    = {North-Holland},
	series       = {Advances in Psychology},
	volume       = 52,
	pages        = {139--183},
	doi          = {https://doi.org/10.1016/S0166-4115(08)62386-9},
	issn         = {0166-4115},
	url          = {https://www.sciencedirect.com/science/article/pii/S0166411508623869},
	editor       = {Peter A. Hancock and Najmedin Meshkati}
}

@article{brooke1996sus,
	title        = {SUS-A quick and dirty usability scale},
	author       = {Brooke, John and others},
	year         = 1996,
	journal      = {Usability evaluation in industry},
	publisher    = {London, England.},
	volume       = 189,
	number       = 194,
	pages        = {4--7}
}

@article{ROSS2025whereto,
	title        = {Forensic science: Where to from Here?},
	author       = {Alastair Ross and Chris Lennard and Claude Roux},
	year         = 2025,
	journal      = {Forensic Science International},
	volume       = 366,
	pages        = 112285,
	doi          = {https://doi.org/10.1016/j.forsciint.2024.112285},
	issn         = {0379-0738},
	url          = {https://www.sciencedirect.com/science/article/pii/S0379073824003670}
}

@article{Hunter2007,
	title        = {Matplotlib: A 2D graphics environment},
	author       = {Hunter, J. D.},
	year         = 2007,
	journal      = {Computing in Science \& Engineering},
	publisher    = {IEEE COMPUTER SOC},
	volume       = 9,
	number       = 3,
	pages        = {90--95},
	doi          = {10.1109/MCSE.2007.55}
}

@article{Waskom2021,
	title        = {seaborn: statistical data visualization},
	author       = {Michael L. Waskom},
	year         = 2021,
	journal      = {Journal of Open Source Software},
	publisher    = {The Open Journal},
	volume       = 6,
	number       = 60,
	pages        = 3021,
	doi          = {10.21105/joss.03021},
	url          = {https://doi.org/10.21105/joss.03021}
}

@online{9news2016,
	title        = {3D reconstruction videos reveal sniper positions during siege stand-off},
	author       = {9NEWS},
	year         = 2016,
	month        = jul,
	day          = 26,
	url          = {https://www.9news.com.au/national/six-more-videos-of-lindt-siege-released},
	note         = {Accessed: 2 April 2025}
}

@article{Whitney1947test,
	title        = {On a Test of Whether one of Two Random Variables is Stochastically Larger than the Other},
	author       = {H. B. Mann and D. R. Whitney},
	year         = 1947,
	journal      = {The Annals of Mathematical Statistics},
	publisher    = {Institute of Mathematical Statistics},
	volume       = 18,
	number       = 1,
	pages        = {50 -- 60},
	doi          = {10.1214/aoms/1177730491},
	url          = {https://doi.org/10.1214/aoms/1177730491}
}

@inbook{Wilcoxon1992,
	title        = {Individual Comparisons by Ranking Methods},
	author       = {Wilcoxon, Frank},
	year         = 1992,
	booktitle    = {Breakthroughs in Statistics: Methodology and Distribution},
	publisher    = {Springer New York},
	address      = {New York, NY},
	pages        = {196--202},
	doi          = {10.1007/978-1-4612-4380-9_16},
	isbn         = {978-1-4612-4380-9},
	url          = {https://doi.org/10.1007/978-1-4612-4380-9_16},
	editor       = {Kotz, Samuel and Johnson, Norman L.}
}

@article{Fei2024gaussiansurvey,
	title        = {3D Gaussian Splatting as New Era: A Survey},
	author       = {Fei, Ben and Xu, Jingyi and Zhang, Rui and Zhou, Qingyuan and Yang, Weidong and He, Ying},
	year         = 2024,
	journal      = {IEEE Transactions on Visualization and Computer Graphics},
	volume       = {},
	number       = {},
	pages        = {1--20},
	doi          = {10.1109/TVCG.2024.3397828}
}

@inproceedings{zheng2023bodytrack,
	title        = {Realistic Full-Body Tracking from Sparse Observations via Joint-Level Modeling},
	author       = {Zheng, Xiaozheng and Su, Zhuo and Wen, Chao and Xue, Zhou and Jin, Xiaojie},
	year         = 2023,
	booktitle    = {2023 IEEE/CVF International Conference on Computer Vision (ICCV)},
	pages        = {14632--14642},
	doi          = {10.1109/ICCV51070.2023.01349}
}

@article{Liu2022posesurvey,
	title        = {Recent Advances of Monocular 2D and 3D Human Pose Estimation: A Deep Learning Perspective},
	author       = {Liu, Wu and Bao, Qian and Sun, Yu and Mei, Tao},
	year         = 2022,
	month        = nov,
	journal      = {ACM Comput. Surv.},
	publisher    = {Association for Computing Machinery},
	address      = {New York, NY, USA},
	volume       = 55,
	number       = 4,
	doi          = {10.1145/3524497},
	issn         = {0360-0300},
	url          = {https://doi.org/10.1145/3524497},
	issue_date   = {April 2023},
	articleno    = 80,
	numpages     = 41
}

@inproceedings{Caine2016samplesize,
	title        = {Local Standards for Sample Size at CHI},
	author       = {Caine, Kelly},
	year         = 2016,
	booktitle    = {Proceedings of the 2016 CHI Conference on Human Factors in Computing Systems},
	location     = {San Jose, California, USA},
	publisher    = {Association for Computing Machinery},
	address      = {New York, NY, USA},
	series       = {CHI '16},
	pages        = {981--992},
	doi          = {10.1145/2858036.2858498},
	isbn         = 9781450333627,
	url          = {https://doi.org/10.1145/2858036.2858498},
	numpages     = 12
}

@inproceedings{Ihara2025Video2MR,
	title        = {Video2MR: Automatically Generating Mixed Reality 3D Instructions by Augmenting Extracted Motion from 2D Videos},
	author       = {Ihara, Keiichi and Monteiro, Kyzyl and Faridan, Mehrad and Kazi, Rubaiat Habib and Suzuki, Ryo},
	year         = 2025,
	booktitle    = {Proceedings of the 30th International Conference on Intelligent User Interfaces},
	location     = {},
	publisher    = {Association for Computing Machinery},
	address      = {New York, NY, USA},
	series       = {IUI '25},
	pages        = {1548--1563},
	doi          = {10.1145/3708359.3712159},
	isbn         = 9798400713064,
	url          = {https://doi.org/10.1145/3708359.3712159},
	numpages     = 16
}

@inproceedings{Pan2020PoseMMR,
	title        = {PoseMMR: A Collaborative Mixed Reality Authoring Tool for Character Animation},
	author       = {Pan, Ye and Mitchell, Kenny},
	year         = 2020,
	booktitle    = {2020 IEEE Conference on Virtual Reality and 3D User Interfaces Abstracts and Workshops (VRW)},
	volume       = {},
	number       = {},
	pages        = {758--759},
	doi          = {10.1109/VRW50115.2020.00230}
}

@inproceedings{Zhang2020Flowmatic,
	title        = {FlowMatic: An Immersive Authoring Tool for Creating Interactive Scenes in Virtual Reality},
	author       = {Zhang, Lei and Oney, Steve},
	year         = 2020,
	booktitle    = {Proceedings of the 33rd Annual ACM Symposium on User Interface Software and Technology},
	location     = {Virtual Event, USA},
	publisher    = {Association for Computing Machinery},
	address      = {New York, NY, USA},
	series       = {UIST '20},
	pages        = {342--353},
	doi          = {10.1145/3379337.3415824},
	isbn         = 9781450375146,
	url          = {https://doi.org/10.1145/3379337.3415824},
	numpages     = 12
}

@article{Bryant2024codesign,
	title        = {Collaborative co-design and evaluation of an immersive virtual reality application prototype for communication rehabilitation (DISCOVR prototype)},
	author       = {Lucy Bryant and Neira Sedlarevic and Peter Stubbs and Benjamin Bailey and Vincent Nguyen and Andrew Bluff and Diana Barnett and Matt Estela and Carolyn Hayes and Chris Jacobs and Ian Kneebone and Cherie Lucas and Poonam Mehta and Emma Power and Bronwyn Hemsley and},
	year         = 2024,
	journal      = {Disability and Rehabilitation: Assistive Technology},
	publisher    = {Taylor \& Francis},
	volume       = 19,
	number       = 1,
	pages        = {90--99},
	doi          = {10.1080/17483107.2022.2063423},
	url          = {https://doi.org/10.1080/17483107.2022.2063423},
	note         = {PMID: 35442823},
	eprint       = {https://doi.org/10.1080/17483107.2022.2063423}
}

@article{Acton2024codesign,
	title        = {Co-design and pilot of a virtual reality intervention to improve mental and physical healthcare accessibility for people with intellectual disability},
	author       = {Acton, Daniel James and Arnold, Rosalyn and Williams, Gavin and NG, Nicky and Mackay, Kirstyn and Jaydeokar, Sujeet},
	year         = 2024,
	month        = jan,
	journal      = {Advances in Mental Health and Intellectual Disabilities},
	volume       = 18,
	number       = 2,
	pages        = {63--75},
	doi          = {10.1108/AMHID-10-2023-0039},
	issn         = {2044-1282},
	url          = {https://doi.org/10.1108/AMHID-10-2023-0039},
	urldate      = {2025-04-08},
	note         = {Publisher: Emerald Publishing Limited}
}

@inproceedings{Shen2023dementia,
	title        = {Dementia Eyes: Co-Design and Evaluation of a Dementia Education Augmented Reality Experience for Medical Workers},
	author       = {Shen, Ximing and Pai, Yun Suen and Kiuchi, Dai and Bao, Kehan and Aoki, Tomomi and Meguro, Hikari and Oishi, Kanoko and Wang, Ziyue and Wakisaka, Sohei and Minamizawa, Kouta},
	year         = 2023,
	location     = {Hamburg, Germany},
	publisher    = {Association for Computing Machinery},
	address      = {New York, NY, USA},
	series       = {CHI '23},
	doi          = {10.1145/3544548.3581009},
	isbn         = 9781450394215,
	url          = {https://doi.org/10.1145/3544548.3581009},
	articleno    = 778,
	numpages     = 18
}

@article{Lantta2024violenceprevent,
	title        = {Co-design of a digital violence prevention and management tool for psychiatric inpatient care: focus on supporting integration into electronic health record system},
	author       = {Lantta, T. and Rautiainen, T. and Anttila, M. and Anttila, J. and Ameel, M.},
	year         = 2024,
	journal      = {European Psychiatry},
	volume       = 67,
	number       = {S1},
	pages        = {S59--S59},
	doi          = {10.1192/j.eurpsy.2024.170}
}

@article{Kosch2023cognitiveloadsurvey,
	title        = {A Survey on Measuring Cognitive Workload in Human-Computer Interaction},
	author       = {Kosch, Thomas and Karolus, Jakob and Zagermann, Johannes and Reiterer, Harald and Schmidt, Albrecht and Wo\'{z}niak, Pawe\l{} W.},
	year         = 2023,
	month        = jul,
	publisher    = {Association for Computing Machinery},
	address      = {New York, NY, USA},
	volume       = 55,
	number       = {13s},
	doi          = {10.1145/3582272},
	issn         = {0360-0300},
	url          = {https://doi.org/10.1145/3582272},
	issue_date   = {December 2023},
	articleno    = 283,
	numpages     = 39
}

@inbook{Wobbrock2016nonparametric,
	title        = {Nonparametric Statistics in Human--Computer Interaction},
	author       = {Wobbrock, Jacob O. and Kay, Matthew},
	year         = 2016,
	booktitle    = {Modern Statistical Methods for HCI},
	publisher    = {Springer International Publishing},
	address      = {Cham},
	pages        = {135--170},
	doi          = {10.1007/978-3-319-26633-6_7},
	isbn         = {978-3-319-26633-6},
	url          = {https://doi.org/10.1007/978-3-319-26633-6_7},
	editor       = {Robertson, Judy and Kaptein, Maurits}
}

@article{BUCK2007dataanim,
	title        = {Application of 3D documentation and geometric reconstruction methods in traffic accident analysis: With high resolution surface scanning, radiological MSCT/MRI scanning and real data based animation},
	author       = {Ursula Buck and Silvio Naether and Marcel Braun and Stephan Bolliger and Hans Friederich and Christian Jackowski and Emin Aghayev and Andreas Christe and Peter Vock and Richard Dirnhofer and Michael J. Thali},
	year         = 2007,
	journal      = {Forensic Science International},
	volume       = 170,
	number       = 1,
	pages        = {20--28},
	doi          = {10.1016/j.forsciint.2006.08.024},
	issn         = {0379-0738},
	url          = {https://www.sciencedirect.com/science/article/pii/S037907380600555X}
}

@article{ctx2264014146750001751,
	title        = {Animated Evidence-Delta 191 Crash Re-Created through Computer Simulations at Trial},
	year         = 1989,
	journal      = {ABA journal.},
	publisher    = {American Bar Association,},
	address      = {Chicago, Ill. :},
	volume       = 75,
	issn         = {0747-0088},
	lccn         = 2009268193
}

@article{schofield2016cglegal,
	title        = {The use of computer generated imagery in legal proceedings},
	author       = {Schofield, Damian},
	year         = 2016,
	journal      = {Digital Evidence \& Elec. Signature L. Rev.},
	publisher    = {HeinOnline},
	volume       = 13,
	pages        = 3,
	doi          = {10.14296/deeslr.v13i0.2293}
}

@article{Galligan2017trajectory,
	title        = {Gunshot wound trajectory analysis using forensic animation to establish relative positions of shooter and victim},
	author       = {Aisling A. Galligan and Craig Fries and Judy Melinek},
	year         = 2017,
	journal      = {Forensic Science International},
	volume       = 271,
	pages        = {e8-e13},
	doi          = {10.1016/j.forsciint.2016.12.039},
	issn         = {0379-0738},
	url          = {https://www.sciencedirect.com/science/article/pii/S0379073817300026}
}

@article{Villa2017humanposeanim,
	title        = {Virtual animation of victim-specific 3D models obtained from CT scans for forensic reconstructions: Living and dead subjects},
	author       = {C. Villa and K.B. Olsen and S.H. Hansen},
	year         = 2017,
	journal      = {Forensic Science International},
	volume       = 278,
	pages        = {e27-e33},
	doi          = {10.1016/j.forsciint.2017.06.033},
	issn         = {0379-0738},
	url          = {https://www.sciencedirect.com/science/article/pii/S0379073817302402}
}

@article{Wu2023ImpersonatAR,
    author = {Wu, Meng-Han and Ipsita, Ananya and Huang, Gaoping and Ramani, Karthik and Quinn, Alex},
    title = {ImpersonatAR: Using Embodied Authoring and Evaluation to Prototype Multi-Scenario Use Cases for Augmented Reality Applications},
    journal = {Journal of Computing and Information Science in Engineering},
    volume = {24},
    number = {3},
    pages = {031007},
    year = {2023},
    month = {10},
    issn = {1530-9827},
    doi = {10.1115/1.4063558},
    url = {https://doi.org/10.1115/1.4063558}
}

@article{Stevenson2024PhoneScanForensic,
author = {Stevenson, Stephanie and Liscio, Eugene},
title = {Assessing iPhone LiDAR \& Recon-3D for determining area of origin in bloodstain pattern analysis},
journal = {Journal of Forensic Sciences},
volume = {69},
number = {3},
pages = {1045-1060},
doi = {https://doi.org/10.1111/1556-4029.15476},
url = {https://onlinelibrary.wiley.com/doi/abs/10.1111/1556-4029.15476},
eprint = {https://onlinelibrary.wiley.com/doi/pdf/10.1111/1556-4029.15476},
year = {2024}
}

@phdthesis{Pooryousef2025Thesis,
author = "Vahid Pooryousef",
title = "{Immersive Forensic Investigation}",
year = "2025",
month = "9",
url = "https://bridges.monash.edu/articles/thesis/Immersive_Forensic_Investigation/30052255",
doi = "10.26180/30052255.v1",
school = "Monash University"
}

@article{Bacchetti2010samplesize,
  title = {Current sample size conventions: Flaws,  harms,  and alternatives},
  volume = {8},
  ISSN = {1741-7015},
  url = {http://dx.doi.org/10.1186/1741-7015-8-17},
  DOI = {10.1186/1741-7015-8-17},
  number = {1},
  journal = {BMC Medicine},
  publisher = {Springer Science and Business Media LLC},
  author = {Bacchetti,  Peter},
  year = {2010},
  month = mar 
}

@Inbook{Braun2022reflexivethematic,
author="Braun, Virginia
and Clarke, Victoria
and Hayfield, Nikki
and Davey, Louise
and Jenkinson, Elizabeth",
editor="Bager-Charleson, Sofie
and McBeath, Alistair",
title="Doing Reflexive Thematic Analysis",
bookTitle="Supporting Research in Counselling and Psychotherapy : Qualitative, Quantitative, and Mixed Methods Research",
year="2022",
publisher="Springer International Publishing",
address="Cham",
pages="19--38",
isbn="978-3-031-13942-0",
doi="10.1007/978-3-031-13942-0_2",
url="https://doi.org/10.1007/978-3-031-13942-0_2"
}

@article{braun2024critical,
  title={A critical review of the reporting of reflexive thematic analysis in Health Promotion International},
  author={Braun, Virginia and Clarke, Victoria},
  journal={Health Promotion International},
  volume={39},
  number={3},
  pages={daae049},
  year={2024},
  publisher={Oxford University Press US}
}

@article{morrison2025forensicscienceISO,
title = {A guide to ISO 21043 Forensic Sciences from the perspective of the forensic-data-science paradigm},
journal = {Science \& Justice},
volume = {65},
number = {5},
pages = {101304},
year = {2025},
issn = {1355-0306},
doi = {https://doi.org/10.1016/j.scijus.2025.101304},
url = {https://www.sciencedirect.com/science/article/pii/S1355030625000887},
author = {Geoffrey Stewart Morrison and Simon Elliott and June Guiness and Lisa Sonden and Denise {Syndercombe Court}}
}

@ARTICLE{maneli2022csrsurvey,
  author={Maneli, Mfundo A. and Isafiade, Omowunmi E.},
  journal={IEEE Access}, 
  title={3D Forensic Crime Scene Reconstruction Involving Immersive Technology: A Systematic Literature Review}, 
  year={2022},
  volume={10},
  number={},
  pages={88821-88857},
  doi={10.1109/ACCESS.2022.3199437}}

@inproceedings{murta1998modelling,
  title={Modelling and rendering for scene of crime reconstruction: A case study},
  author={Murta, Alan and Gibson, Simon and Howard, TLJ and Hubbold, RJ and West, AJ},
  booktitle={Proceedings Eurographics UK},
  pages={169--173},
  year={1998},
  organization={Citeseer}
}

@article{BAILENSON2006,
author = {BAILENSON, JEREMY N. and BLASCOVICH, JIM and BEALL, ANDREW C. and NOVECK, BETH},
title = {Courtroom Applications of Virtual Environments, Immersive Virtual Environments, and Collaborative Virtual Environments},
journal = {Law \& Policy},
volume = {28},
number = {2},
pages = {249-270},
doi = {https://doi.org/10.1111/j.1467-9930.2006.00226.x},
url = {https://onlinelibrary.wiley.com/doi/abs/10.1111/j.1467-9930.2006.00226.x},
eprint = {https://onlinelibrary.wiley.com/doi/pdf/10.1111/j.1467-9930.2006.00226.x},
year = {2006}
}

@article{schofield2009animating,
  title={Animating evidence: computer game technology in the courtroom},
  author={Schofield, Damian},
  journal={Journal of Information, Law and Technology},
  volume={1},
  pages={1--21},
  year={2009},
  publisher={The Electronic Law Journals Project}
}

@ARTICLE{Butler2015-ij,
  title     = "The future of forensic {DNA} analysis",
  author    = "Butler, John M",
  journal   = "Philos. Trans. R. Soc. Lond. B Biol. Sci.",
  publisher = "The Royal Society",
  volume    =  370,
  number    =  1674,
  pages     = "20140252",
  month     =  aug,
  year      =  2015,
  language  = "en"
}

@INPROCEEDINGS{Kudonu2023lasermeasure,
  author={Kudonu, Moza and Koteich, Mohamad Ali and Philip, Sharon Ann and Khokhar, Gursirat and Joseph, Aby and Singh, Nrashant},
  booktitle={2023 International Conference on Computational Intelligence and Knowledge Economy (ICCIKE)}, 
  title={Application of Laser Measurement Device for Crime Scene Measurements}, 
  year={2023},
  volume={},
  number={},
  pages={381-385},
  doi={10.1109/ICCIKE58312.2023.10131819}}

@book{lucy2013introstatforensic,
  title={Introduction to statistics for forensic scientists},
  author={Lucy, David},
  year={2013},
  publisher={John Wiley \& Sons}
}

@article{Balbudhe2025CSRreview,
  title = {Crime scene reconstruction: a scoping review},
  volume = {15},
  ISSN = {2090-5939},
  url = {http://dx.doi.org/10.1186/s41935-025-00495-5},
  DOI = {10.1186/s41935-025-00495-5},
  number = {1},
  journal = {Egyptian Journal of Forensic Sciences},
  publisher = {Springer Science and Business Media LLC},
  author = {Balbudhe,  Mayur Sudhir and Shetty,  B Suresh Kumar and Arun Kumar,  Nayanatara and Singh,  Divyani and Sherekar,  Ayush Ishwar},
  year = {2025},
  month = oct 
}

@article{lee2022hfjurors,
author = {Lee J. Curley and James Munro and Itiel E. Dror},
title ={Cognitive and human factors in legal layperson decision making: Sources 
of bias in juror decision making},
journal = {Medicine, Science and the Law},
volume = {62},
number = {3},
pages = {206-215},
year = {2022},
doi = {10.1177/00258024221080655},
note ={PMID: 35175157},
URL = {https://doi.org/10.1177/00258024221080655},
eprint = {https://doi.org/10.1177/00258024221080655}
}

@article{SMIT2018misleadevidence,
title = {A systematic analysis of misleading evidence in unsafe rulings in England and Wales},
journal = {Science \& Justice},
volume = {58},
number = {2},
pages = {128-137},
year = {2018},
issn = {1355-0306},
doi = {https://doi.org/10.1016/j.scijus.2017.09.005},
url = {https://www.sciencedirect.com/science/article/pii/S1355030617301144},
author = {Nadine M. Smit and Ruth M. Morgan and David A. Lagnado}
}

@article{kaye2009falsepersuasive,
author= {Kaye, David H. },
title={"False, but Highly Persuasive": How Wrong Were the Probability Estimates in McDaniel v. Brown? Commentaries},
journal={Michigan Law Review First Impressions},
volume={108},
pages={1},
year={2009-2010}
}

@article{breuninger1995crime,
  title={Crime Scene Reconstruction Using 3D Computer-Aided Drafting},
  author={Breuninger, PAUL G},
  journal={Police Chief},
  volume={62},
  pages={61--61},
  year={1995},
  publisher={INTERNATIONAL ASSOCIATION OF CHIEFS OF POLICE}
}

@inbook{Dixit2019photogram3Dscan,
	title        = {Which Tool Is Best: 3D Scanning or Photogrammetry -- It Depends on the Task},
	author       = {Dixit, Ishan and Kennedy, Samantha and Piemontesi, Joshua and Kennedy, Bruce and Krebs, Claudia},
	year         = 2019,
	booktitle    = {Biomedical Visualisation : Volume 1},
	publisher    = {Springer International Publishing},
	address      = {Cham},
	pages        = {107--119},
	doi          = {10.1007/978-3-030-06070-1_9},
	isbn         = {978-3-030-06070-1},
	url          = {https://doi.org/10.1007/978-3-030-06070-1_9},
	editor       = {Rea, Paul M.}
}

@article{Cunha2022laserphotocomp,
	title        = {Laser scanner and drone photogrammetry: A statistical comparison between 3-dimensional models and its impacts on outdoor crime scene registration},
	author       = {Rafael Rodrigues Cunha and Claude Thiago Arrabal and Marcelo Mour\~{a}o Dantas and H\'{e}lio Rodrigues Bassanelli},
	year         = 2022,
	journal      = {Forensic Science International},
	volume       = 330,
	pages        = 111100,
	doi          = {10.1016/j.forsciint.2021.111100},
	issn         = {0379-0738},
	url          = {https://www.sciencedirect.com/science/article/pii/S0379073821004205}
}

@inbook{Villa2019photogrammetry,
	title        = {The Application of Photogrammetry for Forensic 3D Recording of Crime Scenes, Evidence and People},
	author       = {Villa, Chiara and Jacobsen, Christina},
	year         = 2019,
	booktitle    = {Essentials of Autopsy Practice: Reviews, Updates and Advances},
	publisher    = {Springer International Publishing},
	address      = {Cham},
	pages        = {1--18},
	doi          = {10.1007/978-3-030-24330-2_1},
	isbn         = {978-3-030-24330-2},
	url          = {https://doi.org/10.1007/978-3-030-24330-2_1},
	editor       = {Rutty, Guy N.}
}

@article{Raneri2018Laserscan,
author = {Domenic Raneri},
title = {Enhancing forensic investigation through the use of modern three-dimensional (3D) imaging technologies for crime scene reconstruction},
journal = {Australian Journal of Forensic Sciences},
volume = {50},
number = {6},
pages = {697--707},
year = {2018},
publisher = {Taylor \& Francis},
doi = {10.1080/00450618.2018.1424245},
URL = {https://doi.org/10.1080/00450618.2018.1424245},
eprint = {https://doi.org/10.1080/00450618.2018.1424245}
}

@article{horvath2024reconstruction,
  title={Reconstruction and Demonstration in Three Dimensions},
  author={Horv{\'a}th, Orsolya Melinda},
  journal={Journal of Eastern European Criminal Law},
  number={02},
  pages={81--94},
  year={2024},
  publisher={Universul Juridic}
}

@article{galves1999not,
  title={Where the not-so-wild things are: computers in the courtroom, the federal rules of evidence, and the need for institutional reform and more judicial acceptance},
  author={Galves, Fred},
  journal={Harv. JL \& Tech.},
  volume={13},
  pages={161},
  year={1999},
  publisher={HeinOnline}
}

@article{BUCK2013recons3d,
title = {Accident or homicide – Virtual crime scene reconstruction using 3D methods},
journal = {Forensic Science International},
volume = {225},
number = {1},
pages = {75-84},
year = {2013},
issn = {0379-0738},
doi = {https://doi.org/10.1016/j.forsciint.2012.05.015},
url = {https://www.sciencedirect.com/science/article/pii/S0379073812002587},
author = {Ursula Buck and Silvio Naether and Beat Räss and Christian Jackowski and Michael J. Thali}
}

@InProceedings{ranglov2024reconsbenefitconcern,
author="Rangelov, Dimitar
and Knotter, Jaap
and Miltchev, Radoslav",
editor="Arai, Kohei",
title="3D Reconstruction in Crime Scenes Investigation: Impacts, Benefits, and Limitations",
booktitle="Intelligent Systems and Applications",
year="2024",
publisher="Springer Nature Switzerland",
address="Cham",
pages="46--64",
isbn="978-3-031-66329-1"
}

@INPROCEEDINGS{jena2020uncertainvissurvey,
  author={Jena, Amit and Engelke, Ulrich and Dwyer, Tim and Raiamanickam, Venkatesh and Paris, Cecile},
  booktitle={2020 IEEE Pacific Visualization Symposium (PacificVis)}, 
  title={Uncertainty Visualisation: An Interactive Visual Survey}, 
  year={2020},
  volume={},
  number={},
  pages={201-205},
  doi={10.1109/PacificVis48177.2020.1014}}

@inproceedings{Brown2004visualvibration,
author = {Brown, Ross},
title = {Animated visual vibrations as an uncertainty visualisation technique},
year = {2004},
isbn = {1581138830},
publisher = {Association for Computing Machinery},
address = {New York, NY, USA},
url = {https://doi.org/10.1145/988834.988849},
doi = {10.1145/988834.988849},
booktitle = {Proceedings of the 2nd International Conference on Computer Graphics and Interactive Techniques in Australasia and South East Asia},
pages = {84–89},
numpages = {6},
location = {Singapore},
series = {GRAPHITE '04}
}

@article{fiedler2003eyedeceiving,
  title={Are your eyes deceiving you: The evidentiary crisis regarding the admissibility of computer generated evidence},
  author={Fiedler, Betsy S},
  journal={NYL Sch. L. Rev.},
  volume={48},
  pages={295},
  year={2003},
  publisher={HeinOnline}
}

\end{document}